\documentclass[Afour,sagev,times]{sagej}

\usepackage{moreverb,url}
\usepackage[colorlinks,bookmarksopen,bookmarksnumbered,citecolor=red,urlcolor=red]{hyperref}
\usepackage{xcolor}
\usepackage{gensymb}
\usepackage{caption}
\usepackage{subcaption}
\usepackage{enumitem}
\usepackage{float}
\usepackage{graphicx}
\usepackage{colortbl}
\usepackage{cleveref}
\usepackage{textcomp}

\newcommand{\figref}[1]{(see Fig. \ref{fig:#1})}

\newcommand{\rpm}{\sbox0{$1$}\sbox2{$\scriptstyle\pm$}
  \raise\dimexpr(\ht0-\ht2)/2\relax\box2 }

\def\volumeyear{2020}

\begin{document}

\runninghead{Rust et al.}

\title{A data acquisition setup for data driven acoustic design}

\author{Romana Rust\affilnum{1},
        Achilleas Xydis\affilnum{1},
        Kurt Heutschi\affilnum{3},
        Nathanael Perraudin\affilnum{2},
        Gonzalo Casas\affilnum{1},
        Chaoyu Du\affilnum{1},
        J\"urgen Strauss,
        Kurt Eggenschwiler\affilnum{3},
        Fernando Perez-Cruz\affilnum{2},
        Fabio Gramazio\affilnum{1} and
        Matthias Kohler\affilnum{1}
}

\affiliation{
\affilnum{1}Gramazio Kohler Research, ETH Zurich, Switzerland\\
\affilnum{2}Swiss Data Science Center, Switzerland\\
\affilnum{3}Laboratory for Acoustics / Noise Control, Empa, Switzerland\\
}

\corrauth{Romana Rust
\\Chair of Architecture and Digital Fabrication\\ETH Zurich, HIB E 43\\Stefano-Franscini-Platz 1, 8093 Zurich, Switzerland
}
\email{rust@arch.ethz.ch}

\begin{abstract}
In this paper, we present a novel interdisciplinary approach to study the relationship between diffusive surface structures and their acoustic performance. Using computational design, surface structures are iteratively generated and 3D printed at 1:10 model scale. They originate from different fabrication typologies and are designed to have acoustic diffusion and absorption effects. An automated robotic process measures the impulse responses of these surfaces by positioning a microphone and a speaker at multiple locations.
The collected data serves two purposes: first, as an exploratory catalogue of different spatio-temporal-acoustic scenarios and second, as data set for predicting the acoustic response of digitally designed surface geometries using machine learning. In this paper, we present the automated data acquisition setup, the data processing and the computational generation of diffusive surface structures. We describe first results of comparative studies of measured surface panels and conclude with steps of future research.
\end{abstract}


\maketitle

\section{Introduction}
\label{intro}

The acoustic quality of a room is an important criterion for the perception and subsequently the sense of well-being for its inhabitants \cite{handelListeningIntroductionPerception1993, yangDesignStrategiesElements2020}. However, today's architectural acoustic design is mainly focused on typologies that demand high-end acoustics, like concert halls or auditoriums. The acoustic design of the vast majority of the built environment is often overlooked, leading to reduced comfort, negative health effects from acoustic pollution, cost for noise abatement measures and unaesthetic retrofitting of built structures both indoors and outdoors. 

One of the main reasons for this is the lack of accurate and easy-to-use simulation tools \cite{pelzerIntegratingRealTimeRoom2014} that can be well integrated into computational design workflows, enabling the assessment of acoustic quality without the need for acoustic specialists. Thus, acoustics is only considered at a later stage of the architectural planning process (and often concerns only the installation of standard absorption panels).
Still, computational room acoustics is a field that has been intensively studied over the past 50 years \cite{saviojaOverviewGeometricalRoom2015a}. Fundamentally, there are two main approaches for computationally modelling the acoustics of a room, which are either based on numerically solving the wave equation, or on the assumptions of geometrical acoustics (GA). Wave-based modeling is able to provide the most accurate results, but is too computationally expensive \cite{coxAcousticAbsorbersDiffusers2004, siltanenRaysWavesUnderstanding2010} for an iterative design and evaluation workflow. GA is faster, but less accurate. Here sound is assumed to propagate as rays and the wave-nature of sound is neglected. Thus, all wave-based phenomena, such as diffraction and interference are missing. Available room acoustic modelling software such as ODEON \cite{naylorODEONAnotherHybrid1993}, CATT, EASE, Ramsete \cite{ramsete_farina} or RAVEN \cite{schroderRAVENRealtimeFramework2011} are offering hybrid GA methods, where the image source approach is combined with ray-tracing that allows to consider diffuse reflections \cite{vorlander_computer_2013}. The scattering properties of a surface are usually described by a simple one-parameter model that assumes Lambert reflection directivity. This approach splits up the reflected power into a specular and a scattering part, whereas the ratio between the two contributions depends on the frequency and the structure depth. This coarse reflection model can not consider specific surface properties that can generate particular reflection patterns. In order to be able to work in room acoustic design with surfaces with specially designed reflective properties, other solutions are necessary. 

Another method to validate room acoustics utilize physical scale models \cite{katzPhilharmonieParisAcoustic2015}. Here, sound sources are installed at pre-defined positions, emitting sound in a scaled frequency range while the corresponding audio signals are recorded. The resulting measurements can be used to analyse the acoustic performance \cite{coxAcousticAbsorbersDiffusers2004} and improve the design \cite{koren_grand_2018}. However, this method is extremely time- and resource-inefficient, as the number of design iterations are limited to the number of built models.

In this paper, we present a novel interdisciplinary approach to study the mutual relationship between diffusive surface textures and their acoustic performance through data science methods. In order to leverage data gathered from physical scale models, we employ an automated robotic measurement setup to record the impulse responses in front of 3D printed acoustically diffusive surfaces at 1:10 scale. They represent surface structures created through certain fabrication typologies, such as brick or stone walls, for which we collect diverse acoustic scenarios. The recorded data set serves as a foundation to analyse relations between geometrical and acoustical configurations and to determine performance clusters.
The final goal is to use the created data set as a basis for a data-driven acoustic simulation that will allow to predict the acoustic properties of newly created 3D surfaces, thus omitting the need for a physical scale model.

The main challenge of building this data set arises from the need to define and collect sufficient, relevant, and reliable data in a short amount of time. Additionally, 
the post-processing of the input data needs to be identified, since both geometric and acoustic information are high dimensional. This is necessary for both the data visualisation and the future ML system.  
In the following sections, we describe the data acquisition setup, the parameters of the data set and the post-processing of the impulse response to extract meaningful measures, such as the reflected cumulative energy per frequency band. These evaluated indicators allow different panels to be compared. We introduce the computational generation of diffusive surface structures and conclude with strategies for shaping the data set and future work. 

\section{Acoustic data acquisition setup}
\label{setup}

The constituent parts of the multi-robotic setup were developed collaboratively by evaluating architectural and acoustic requirements, in addition to the requirements from the perspective of data science and the constraints of a physical setup. Several tests were performed to guide the development and to validate the quality of the measured data. Some of these tests can be found in the project's open data repository \cite{data-driven-acoustic-design}. 

\begin{figure}[htb]
\centering
\includegraphics[width=\linewidth]{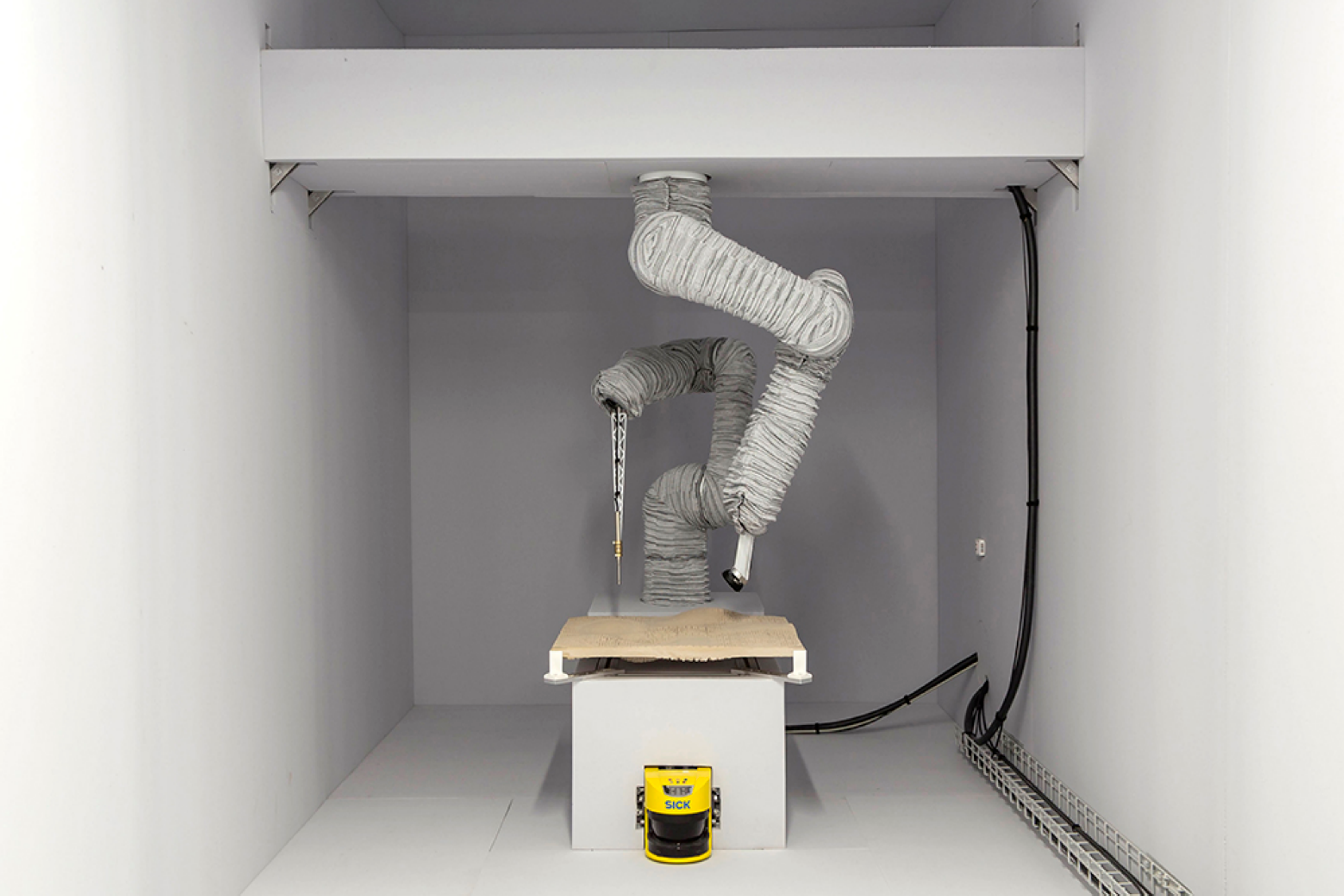}
\captionsetup{width=\linewidth}
\caption{Acoustic data acquisition setup with two Staubli TX2-60L robots in an acoustically shielded and absorbent room.}
\label{fig:setup}
\end{figure}

\begin{figure}[htb]
\centering
\captionsetup{width=\linewidth}
\includegraphics[width=\linewidth]{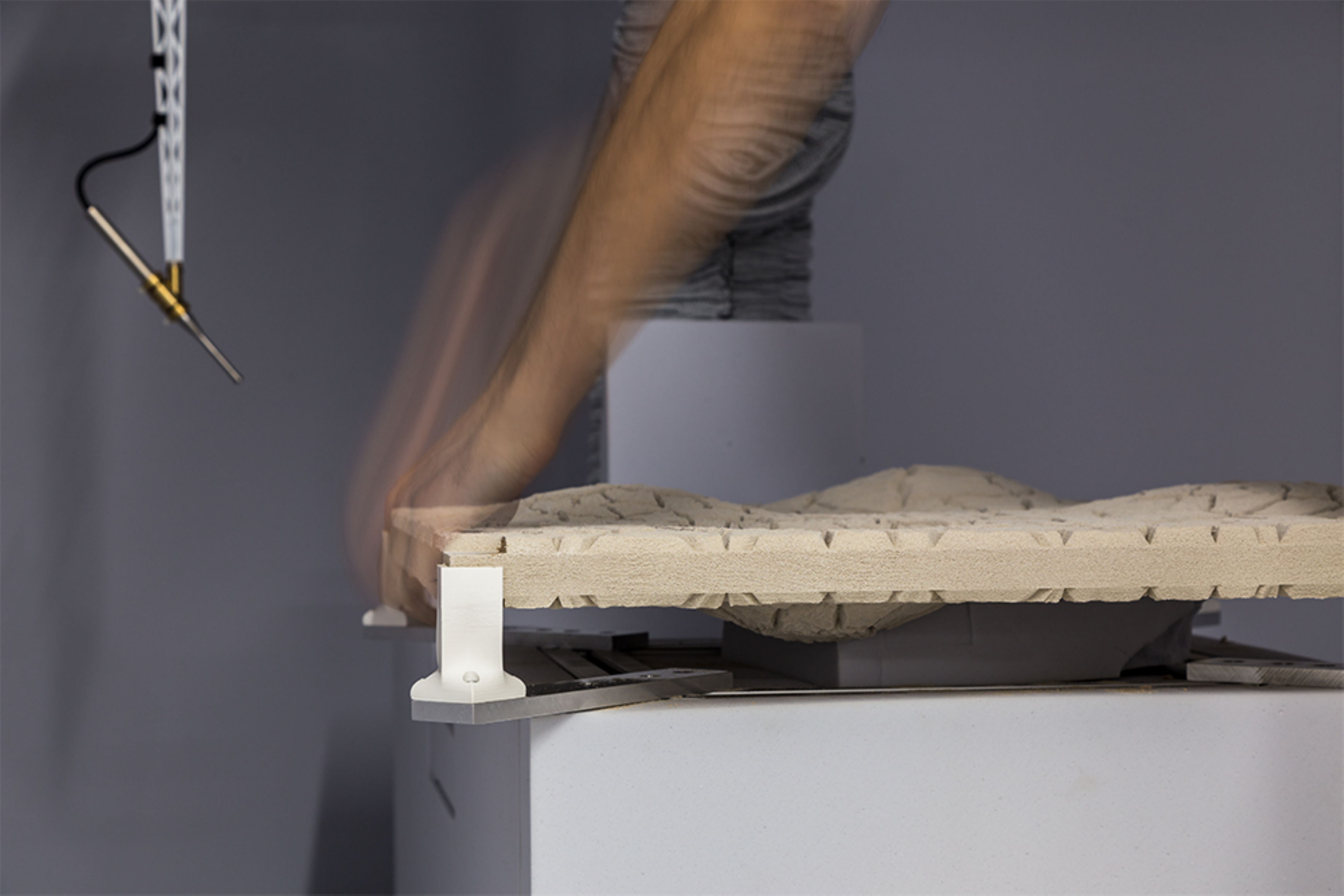}
\caption[setup]{Double-sided 3D printed panel placed in a special fixture.}
\label{fig:doublesidedpanel}
\end{figure}

The multi-robotic measurement setup consists of two 6-axis Staubli TX2-60L\footnote{These robotic arms are accurate (absolute positioning accuracy 0.2 mm, repeatability 0.02 mm) and they have the ability to programmatically turn the joint motors off and on, such that their operating noises do not affect acoustic measurements.} robotic arms with a reach of 920 mm each \figref{setup}. They are equipped with two different end-effectors: one with a speaker and the other one with a microphone. During the measurement process they reconfigure from position to position in an irregular measurement grid above a 3D printed acoustic panel. For each combination of microphone and speaker position, a sweep signal in a scaled (1:10) frequency range of 2--40 kHz is emitted, a recording is taken, and the corresponding impulse response (IR) is calculated. The sweep signal covers the frequency range that can be reproduced by the loudspeaker and determines the lower and upper frequency limit of the data. The time spent on each measurement combination averages at 12.3 seconds and the measurement process per acoustic surface takes approximately 10 hours, during which the data of 2951 measurement combinations are stored. To avoid acoustic reflections, the robotic arms are covered with custom 3D knitted sound absorbing cloths. The robot controllers are installed in the adjacent room to prevent their operating noise from affecting the measurement. The whole setup is installed inside a sound insulated room, in which all surfaces are covered with 50 mm melamine foam\footnote{Basotect\textsuperscript{\textregistered} G+ Melamine foam from Vibraplast AG.}. In the following paragraphs, the core components of the setup are described.

\subsection{3D printed acoustic panels}
The goal of the research project is to produce a large and rich data set during the project's time span. However, the main constraints are the measurement time, the acoustic panel's size and its fabrication time. The print-bed of the in-house Voxeljet VX1000 3D sand printer and the defined operation hours, constrain the acoustic panel's size to a bounding box measuring 585x585x100 mm (WxLxH), enabling the production of maximal five panels per week. 
To increase the amount of measurable surfaces, we designed the panel with two sides \figref{doublesidedpanel}, thus two acoustic surfaces per panel. A square panel shape was selected for the possibility of applying standard data augmentation techniques: based on the assumption that the measurements are symmetric, the data can be virtually mirrored and rotated four times (90 degrees each time), resulting in an overall increase of the collected data by a factor of eight. 

A panel produced with a binder-jet 3D printer is porous and highly absorbing. In order to obtain a surface that represents rigid non porous materials, the panel is coated. We evaluated different surface treatments and compared the respective normal incidence absorption coefficients obtained by impedance tube measurements. If left untreated, or baked in an oven, the absorption coefficient is 0.47-0.58 for frequencies between 2--6 kHz. If infiltrated with resin, or coated with two layers of acrylic paint, the absorption coefficient is below 0.1. 
We decided to proceed with the application of two layers of a plant-based, water-borne paint using a compressed air spray gun. We compared the panel's surface reflectivity after coating by comparing the measurements from a coated flat 3D printed surface (referred to as \emph{Flat}, see Table \ref{tab:baseline}) with a flat MDF panel (referred to as \emph{Wood}). 
Compared to \emph{Wood}, \emph{Flat} reflected on average 29.3\% less energy. This unavoidable loss in reflected energy and the variations of the measurements is considered in the subsequent evaluations by normalisation (see Section \nameref{sec:IR}), that is to say all indicators are consequently calculated in relation to \emph{Flat}.

\subsection{Measurement grid} 
\begin{figure}[htb]
\centering
\captionsetup{width=\linewidth}
\includegraphics[width=\linewidth]{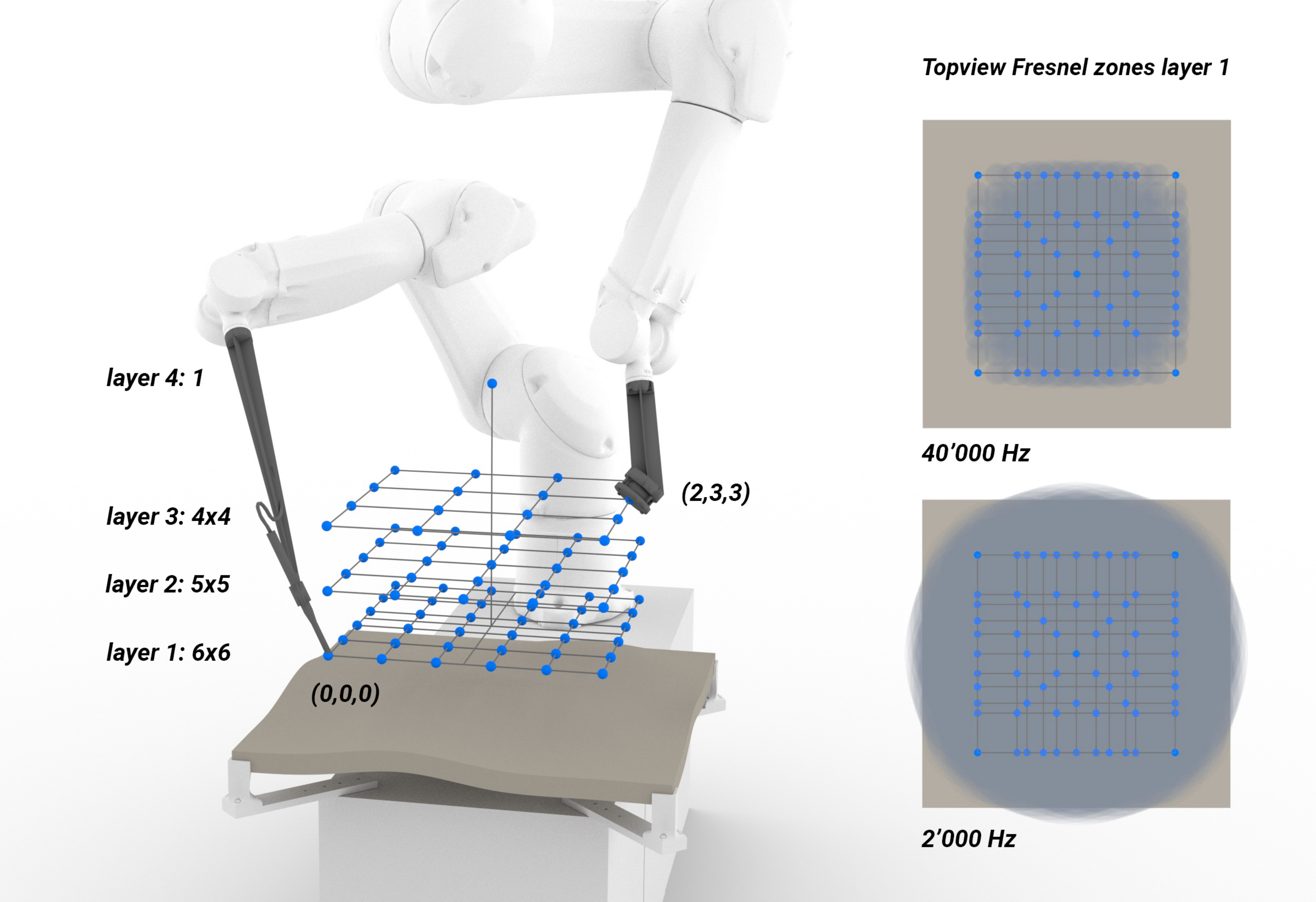}
\caption[setup]{Left: Measurement grid with 4 layers in relation to robotic setup and 3D printed panel. Right: Panel topview with surface coverage in layer 1 at 40'000 Hz (top) and 2'000 Hz (bottom) calculated by Fresnel zones.}
\label{fig:gridfresnel}
\end{figure}

The measurement locations are set in an irregular point grid based on the defined dimensions of the 3D printed panel, the robots' working space, and acoustic considerations. The grid’s dimensions are defined to avoid measurements with edge diffraction as much as possible. The density of the measurement grid was calculated based on three criteria: a) to ensure a uniform surface coverage, b) to maximize the number of data points per panel, and c) to allow two acoustic surfaces to be measured within a 24-hour cycle. To do so, we calculated the first Fresnel zone\cite{heddle_fresnel_2016} \footnote{Fresnel zones on a surface are the intersections of Fresnel spheroids with a flat surface between a source and the image of the receiver. The foci of the Fresnel spheroid are the source and the image of the receiver. The resulting intersections have the form on an ellipse.} for each microphone and speaker combination for both the lowest and highest used frequencies, assuming a planar surface. By calculating all possible combinations of speaker and microphone positions (excluding some immeasurable cases), the final measurement grid contains 78 measurement points (see Fig. \ref{fig:gridfresnel}) and a total of 2951 measurement combinations. The measurement points are placed on four planar layers, each with a different number of measurement points located at different offsets from the panel's surface. The first layer contains 6x6 measurement points, the second 5x5, and the third 4x4, with average offsets of 124, 214, and 304 mm from the surface and respective distances of 75, 93.75 and 125 mm between measurement points. The fourth layer has only one measurement point with an average offset of 474 mm from the surface. Finally, the Fresnel zones for the lowest frequency (2 kHz) have a minimum ellipse diameter of 195 mm and a maximum of 560 mm, and for the highest frequency (40 kHz) 43 mm and 140 mm respectively.

\begin{figure}[htb]
\centering
    \captionsetup{width=\linewidth}
    \includegraphics[width=\linewidth]{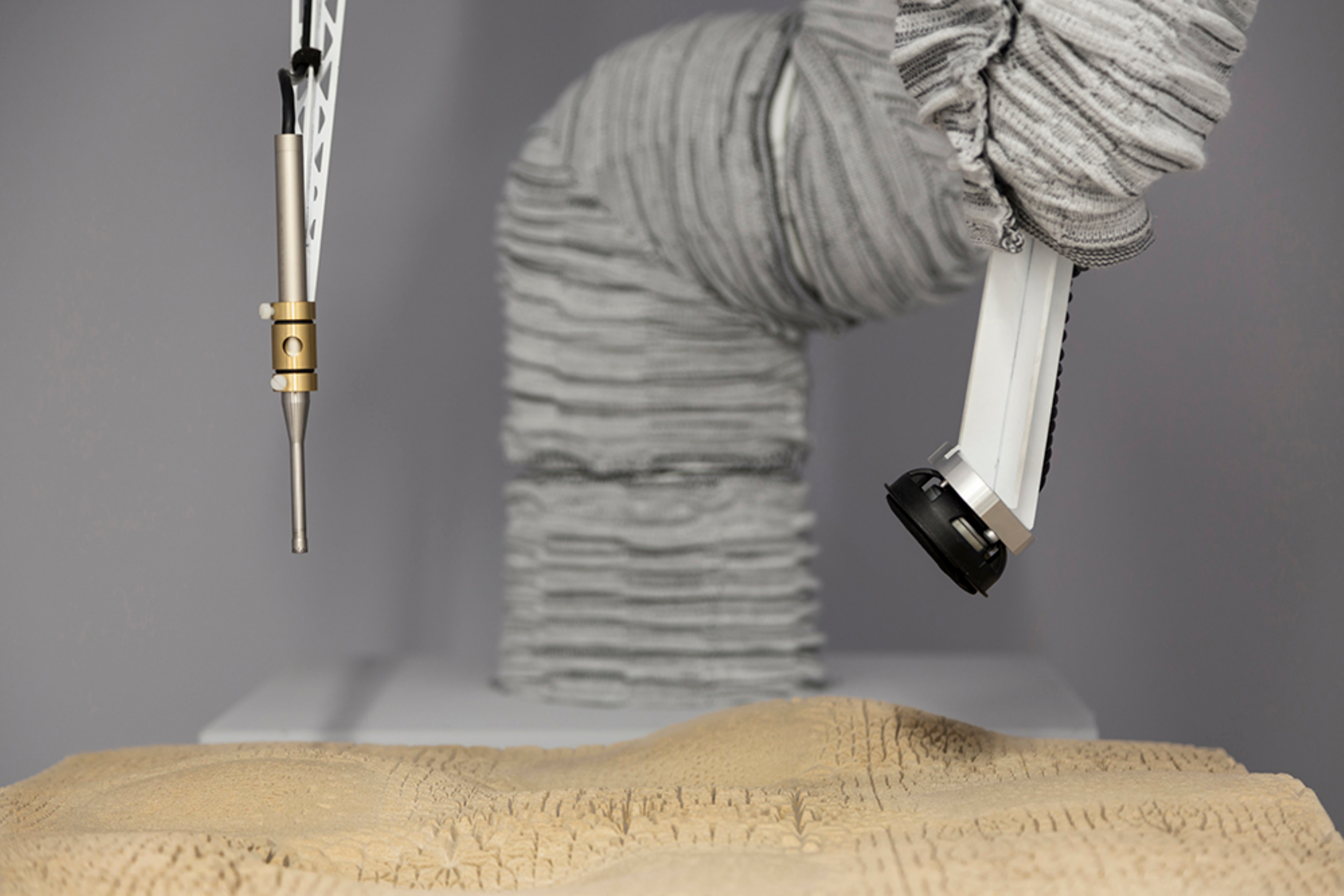}
    \caption[setup]{Microphone (left) and speaker end-effector (right). The microphone is attached to an acoustically transparent steel mount and the speaker is tilted to optimize directivity and ensure robot reachability.}
    \label{fig:tools}
\end{figure}

\begin{table*}[!htb]
\centering
\begin{tabular}{l l p{0.45\textwidth}}
\toprule
Label & Material & Purpose\\
\midrule 
\emph{Wood} & MDF plate & Reference for a surface of high reflection.\\
\hline
\emph{Flat} & 3D printed and coated &  Reference for the surface of highest reflection possible with the used 3D printed and coated material. Used to normalise the measurements.\\
\hline
\emph{Foam} & Acoustically absorbent melamine foam\footnote{Basotect\textregistered G+ melamine foam from Vibraplast AG. The same foam was used for the surfaces of the setup room.} & Reference for a surface of high absorption. Used for subtracting the direct sound signal from the measurement.\\
\hline
\emph{2D-PRD} & 3D printed and coated & 2D Primitive Root Diffuser. Reference for a surface of high and uniform diffusion.\\
\hline
\emph{0015\_0} & 3D printed and coated & Reference for a specific macrostructure with no meso- and microstructure.\\
\bottomrule
\end{tabular}
\captionsetup{justification=centering}
\caption{Reference panels}
\label{tab:baseline}
\end{table*}

\subsection{Microphone and speaker end-effector} 
In order to record clean audio responses, shielding and scattering from the robotic arms is avoided to the greatest possible extent. The microphone is positioned such that it is far from the robot's flange (approx. 0.5 m) and it is fixed on an acoustically transparent steel mount \figref{tools}. The precise tool manufacturing and the accuracy of the robotic arm allows to achieve a positional accuracy of 0.17 mm\footnote{After the absolute calibration of the robotic arms we have a mean precision of 0.17 mm and a max@90\% of 0.28 mm.} for the microphone. The microphone consists of a G.R.A.S. 40BE capsule attached on a Microtech Gefell MV 220 high-impedance transducer. The microphone is of free-field type and has a flat amplitude response up to 40 kHz (-1 dB) for sound incidence on axis. For 30$^{\circ}$ and 60$^{\circ}$ off-axis, the sensitivity at 40 kHz drops by 2 and 4 dB respectively. 

On the source side, a Beryllium tweeter was selected as a loudspeaker that is capable of exiting frequencies between 2 and 40 kHz. As a direct consequence of the 20 mm membrane diameter, the loudspeaker shows a directivity pattern with a tendency to focus sound radiation on axis at high frequencies. Several tests with conical attachments and scattering objects in front of the membrane showed an improved (closer to omni-directional) radiation pattern, however with a degradation of the temporal signature. To maintain the excellent time response of the tweeter, it was decided to do without measures to optimize directivity but carefully orient the speaker in each measurement configuration. This is the reason for the 45$^{\circ}$ tilt of the steel mounting \figref{tools}, ensuring reachability by the robotic arm.

The microphone and speaker are connected to a Focusrite Scarlett 2i4 2nd Gen audio interface. We use two of the mono balanced output channels. One is connected to an amplifier that drives the Beryllium tweeter and the other is connected back to one of the audio interface's inputs and used as a loopback channel for computing the impulse response (IR) (See \nameref{sec:IR}).

\subsection{Automation, control setup and sensors} 
COMPAS FAB\cite{compas-fab2018} and MoveIt\cite{colemanMoveIt2014} were used to calculate collision-free robot trajectories for each of the 2951 measurement configurations of microphone and speaker along a defined sequence. For the data acquisition phase, a workstation running Ubuntu 16.04, together with the audio interface and the two Stäubli CS9 robot controllers were installed in the adjacent control room. ROS Kinetic\cite{quigleyRos2009} is used as the base of a distributed system with the following nodes: main controller service, ambient measurement service, audio interface service, two VAL3 robot driver instances and websockets ROS bridge\cite{torisRosBridge2015}. The main controller service was built using COMPAS FAB\cite{compas-fab2018} and it coordinates all other services. After positioning, the controller powers the robots off, so that their operating noises do not affect the measurement. Then it invokes the audio interface to start playback and recording while the ambient measurement service collects external sound level, temperature, relative humidity, and atmospheric pressure using an Arduino board. The external sound level values are employed to track exogenous sounds that can influence the quality of our measurements. After recording, the impulse response is calculated and validated to ensure that the measurement is not distorted by unwanted signals, and, repeated if needed.

Metrics of the process are continuously collected in an \textit{InfluxDB} time-series database and \textit{Grafana} is used for monitoring. Tracked metrics include values from all ambient measurement sensors, system metrics based on \textit{collectd}, and process metrics.

\begin{figure*}[!htb]
\captionsetup[subfigure]{justification=centering}
\centering
    \begin{subfigure}[t]{.32\textwidth}
        \includegraphics[width=\textwidth]{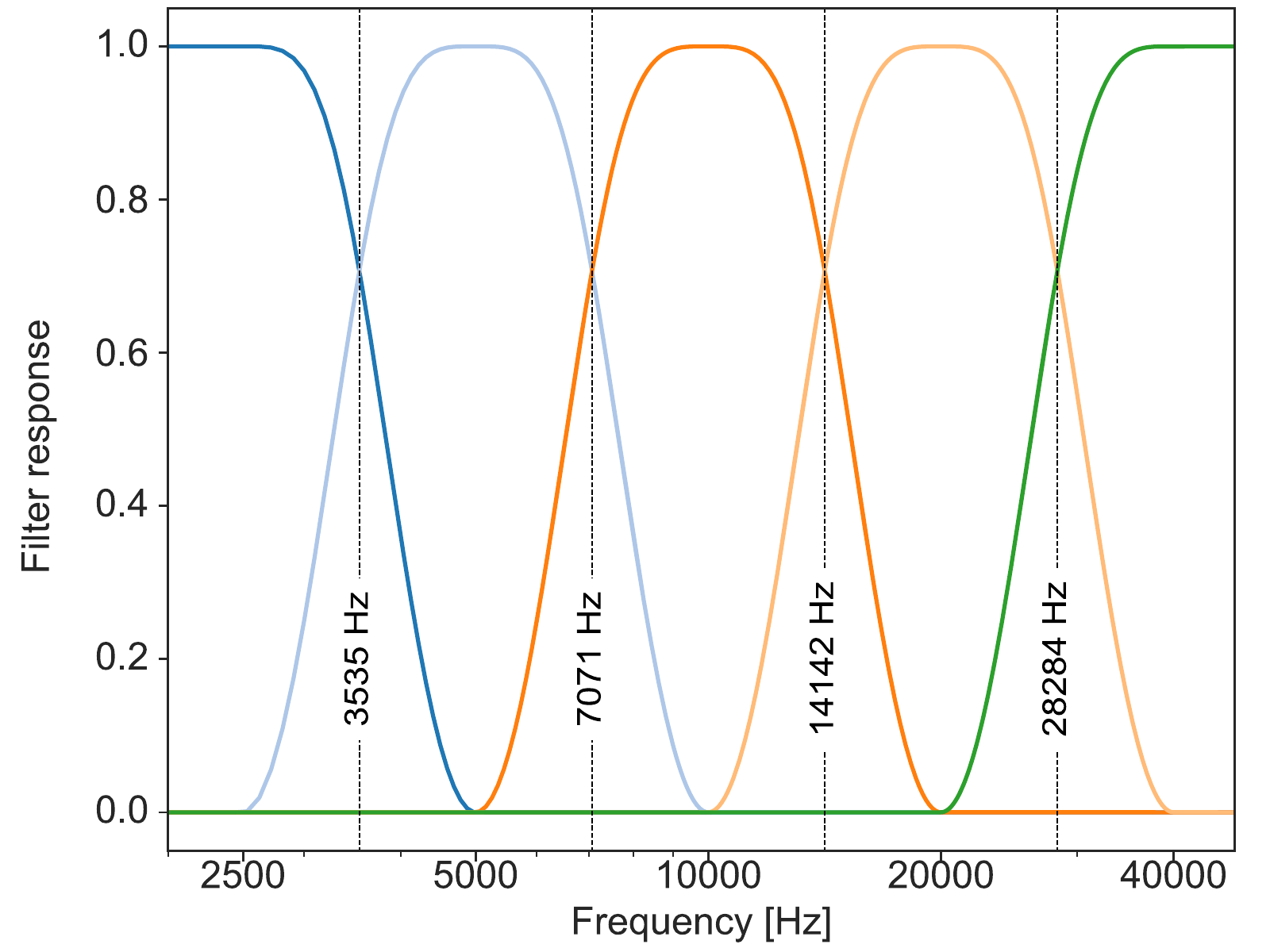}
    	\caption{Filter bands}
    	\label{fig:filterbank}		
	\end{subfigure}
	\begin{subfigure}[t]{.32\textwidth}
        \includegraphics[width=\textwidth]{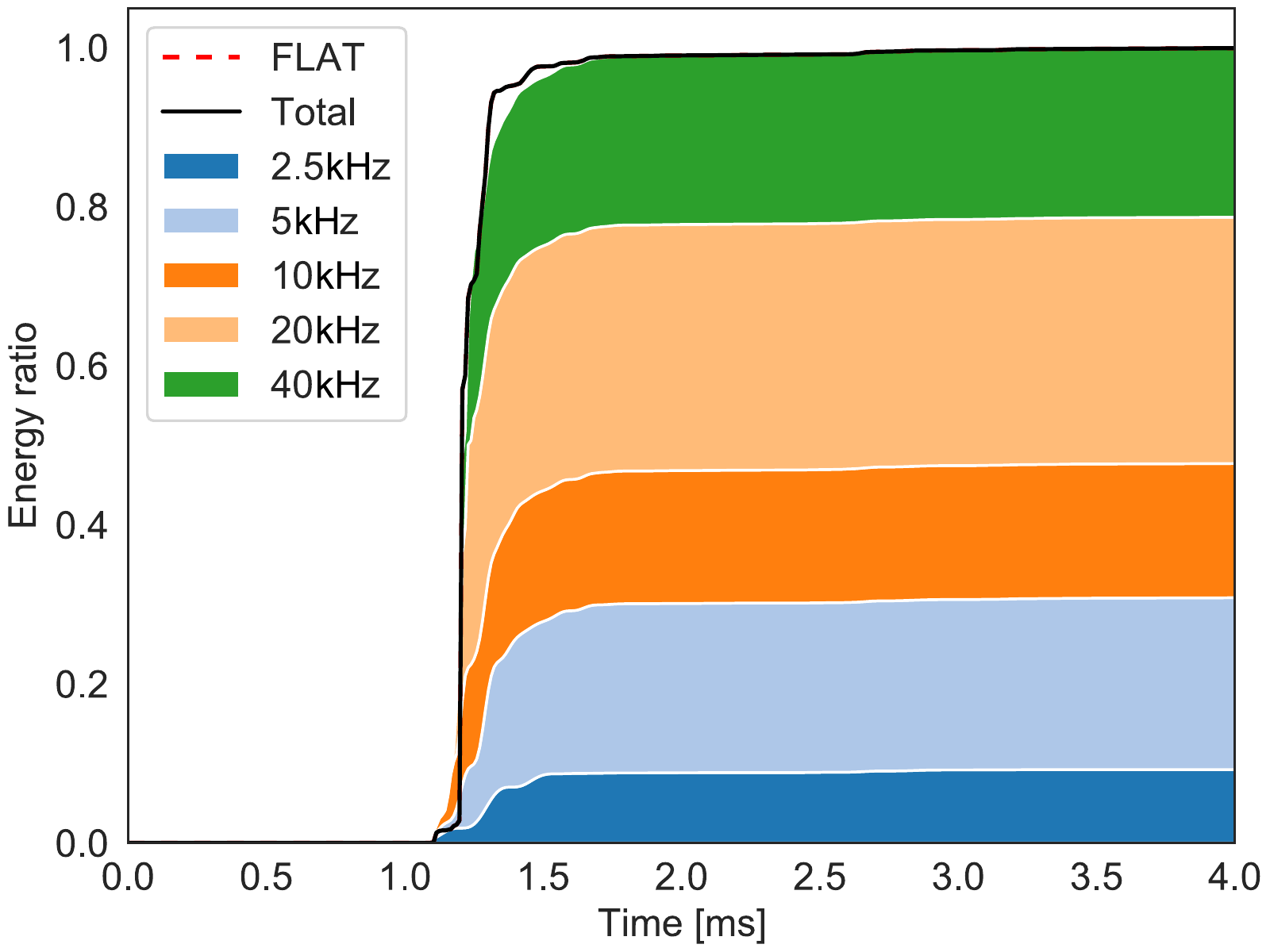}
		\caption{Cumulative energy of \emph{Flat}}
		\label{fig:cum_energy_plot_flat}
	\end{subfigure}
	\begin{subfigure}[t]{.32\textwidth}
        \includegraphics[width=\textwidth]{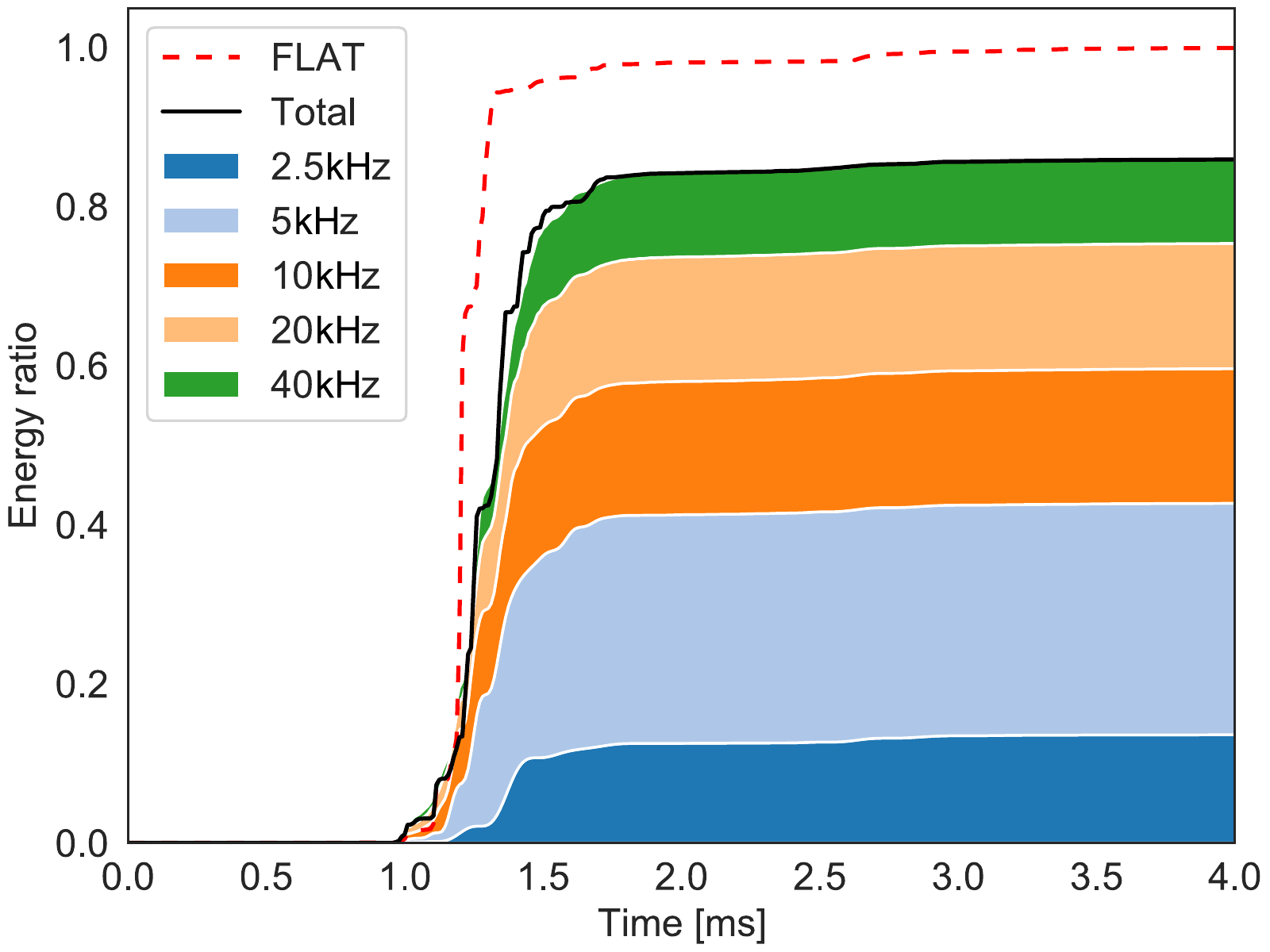}
		\caption{Cumulative energy of panel \emph{0072\_0}}
		\label{fig:cum_energy_plot_0072}
	\end{subfigure}
\caption{(\ref{fig:filterbank}) Constructed filters to separate the content of the IR in frequency. Stacked cumulative energy curves of panel \emph{Flat} (\ref{fig:cum_energy_plot_flat}) and panel \emph{0072\_0} (\ref{fig:cum_energy_plot_0072}) . The colours relate to the filter band colours of Fig. \ref{fig:filterbank}.}
\label{fig:data-post-process}
\end{figure*}

\section{Data set, post-processing and visualisation}

The acoustic data acquisition setup collects different spatio-temporal-acoustic scenarios, which are stored in a multivariable data set. 
One data point in the data set consists of the computationally generated geometry of the measured diffusive surface, plus 2951 impulse responses, supplemented by measured environmental data (temperature, humidity, atmospheric pressure). 
The geometric information of the data set includes input parameters of the geometry generation algorithm (see \nameref{design}), together with the algorithm itself, and the representation of the surface as a polygon mesh. The mesh data is directly used for the panel fabrication with the binder-jet 3D printer. Additionally, each 3D panel is labeled with a unique identifier and suffixed with 0 or 1 indicating the panel side (e.g. \emph{0015\_0}). This identifier is used to determine the 3D printed physical object with the data set entry.

\subsection{Reference panels}
Some datapoints in the data set are baseline measurements obtained from special reference panels with the same dimensions as our 3D printed acoustic surfaces. These serve to put the measurements of the 3D printed acoustic panels in relation to other materials or panels with a different surface geometry. Table \ref{tab:baseline} lists the baseline measurements with their respective label, material and purpose. Two of these baseline panels (\emph{Flat} and \emph{Foam}) are also used in the post-processing of the data, which is part of both the ML processing pipeline and the data visualisation.

\subsection{Impulse response and data post-processing}
\label{sec:IR}
The primary measurement result for a specific surface and speaker/microphone combination is the impulse response (IR).
The IR is the richest representation possible as it contains all of the acoustic information linking the source and the receiver. Furthermore, one advantage of the IR is the fact that different contributions appear lined up on the time axis. 
As a result, the 2951 IRs offer a very precise and relatively complete representation of the acoustic response of a panel surface. Nevertheless, the IRs present some challenges as well. First, IRs are not easily interpretable with respect to perceptional aspects, especially because the phase information is very complex. Second, for human data analysis it is necessary to compress the information contained in the 2951 IRs such that it can be comprehensibly visualised per acoustic surface. Third, for the future ML system the direct modelling of the IRs might be challenging or even impossible given the low amount of available samples at the end of the research project. In consequence, we identified other indicators that represent the desired acoustic information of a surface from an acoustic design perspective.

\subsubsection{Information extraction.}
First, to obtain the IR, we play a linear frequency sweep ranging from 2 to 40 kHz and record the microphone signal as raw data. The IR is then computed by deconvolution and temperature compensation is applied. Afterwards, we crop the IR after 4 ms and suppress the direct sound. Due to small path length differences between direct and reflected sound in some geometries, a time-windowing based separation is not applicable. For that reason, the direct sound time signal obtained from an IR measurement with an absorbing panel (referred to as \emph{Foam}) is subtracted.  
Third, the IR is band-pass filtered with the help of the filter bank described below \figref{filterbank}. This allows the derivation of frequency dependent reflection properties of the surface. Details about these first three steps are available in Appendix~\nameref{appendix:impulse_response}.
Fourth, the filtered IRs are converted to cumulative energy curves \figref{cum_energy_plot_flat} that display, on one hand, total reflected energy and its distribution among the different filter bands, and, on the other hand, the temporal pattern of energy arrival. The cumulative energy curves are then put in relation to the measurements obtained from a reference flat panel (referred to as \emph{Flat}) by normalisation. For simplification, we refer to the resulting curves as \emph{NCE} curves and the resulting total value as \emph{TNCE} in the following. The \emph{TNCE} measure allows comparing different panels with each other. For example, if we contrast the stacked cumulative energy plots of Fig. \ref{fig:cum_energy_plot_flat} and Fig. \ref{fig:cum_energy_plot_0072} and refer to Table \ref{table:cum_energy_flat_0072} for the \emph{TNCE} values, the following information can be extracted: First, we see that panel \emph{0072\_0} reflects 14.1\% less energy than the \emph{Flat} panel. Second, the energy distribution among the different frequency bands changes. The 2.5 kHz and 5 kHz bands are exhibiting an energy increase, 10 kHz band has almost no difference ($<$0.2\%), and the two higher bands a significant decrease. Additionally, the slope of the \emph{NCE} curve relates to the degree of diffusiveness where a steep gradient indicates a rather specular reflection and a slow increase represents a diffuse reflection. In this case, panel \emph{0072\_0} has a slightly less steep slope.

\begin{table}[htb]
\centering
\resizebox{\linewidth}{!}{
\begin{tabular}{c c c c c c c c } 
\hline
Panel ID & 2.5 kHz & 5 kHz & 10 kHz & 20 kHz & 40 kHz & Total \\ 
\hline\hline
Flat & 0.092 & 0.216 & 0.169 & 0.309 & 0.213 & 1.000 \\ 
\hline
0072\_0 & 0.136 & 0.291 & 0.169 & 0.157 & 0.106 & 0.859 \\
\hline\hline
energy & 47.4\% & 34.5\% & 0.18\% & -49.2\% & -50.3\% & -14.1\% \\
difference & \multicolumn{1}{l}{{\cellcolor[rgb]{0.125,0.463,0.702}}} & \multicolumn{1}{l}{{\cellcolor[rgb]{0.682,0.78,0.91}}} & \multicolumn{1}{l}{{\cellcolor[rgb]{1,0.498,0.059}}} & \multicolumn{1}{l}{{\cellcolor[rgb]{1,0.733,0.471}}} & \multicolumn{1}{l}{{\cellcolor[rgb]{0.169,0.627,0.176}}} & \multicolumn{1}{l}{{\cellcolor[rgb]{0,0,0}}} \\
\hline
\end{tabular}
}
\captionsetup{justification=centering}
\caption{\emph{TNCE} values for \emph{Flat} and panel \emph{0072\_0}. The values relate to Fig. \ref{fig:cum_energy_plot_flat} and Fig. \ref{fig:cum_energy_plot_0072}}
\label{table:cum_energy_flat_0072}
\end{table}


\subsubsection{Filterbank design.}
The signal is separated into different frequency bands using an "itersine" wavelet construction. Formally, we use the mother function $c(\omega) = \sin \left(\frac{\pi}{2} \cos(\pi \omega) \right)$ and scale, warp and translate it as in \cite{holighaus2020class}. Selecting the right parameters, we construct the set of five filters shown in  \figref{filterbank}. The filters are centered at 5, 10, 20 kHz and are logarithmically stretched (warped). The blue and green filters correspond to the the remaining low and high frequency bands. Note that, because everything is 1:10 scaled, the three bandpass filters correspond to 0.5, 1 and 2 kHz bands. Note that this set of filters form a unitary tight frame, meaning that the total energy of the signal is conserved after the application of the filters.
The proposed construction does not satisfy a particular norm for octave based filter bank such as IEC 61260.1:2019 \cite{iec1995electroacoustics}. However, it is tailored to our application because it conserves the energy and has good localization properties both in the time and the frequency domain.

\subsection{Data visualisation}
\begin{figure}[htb]
\captionsetup[subfigure]{justification=centering}
\centering
    \begin{subfigure}[t]{.48\textwidth}
        \sbox0{\includegraphics{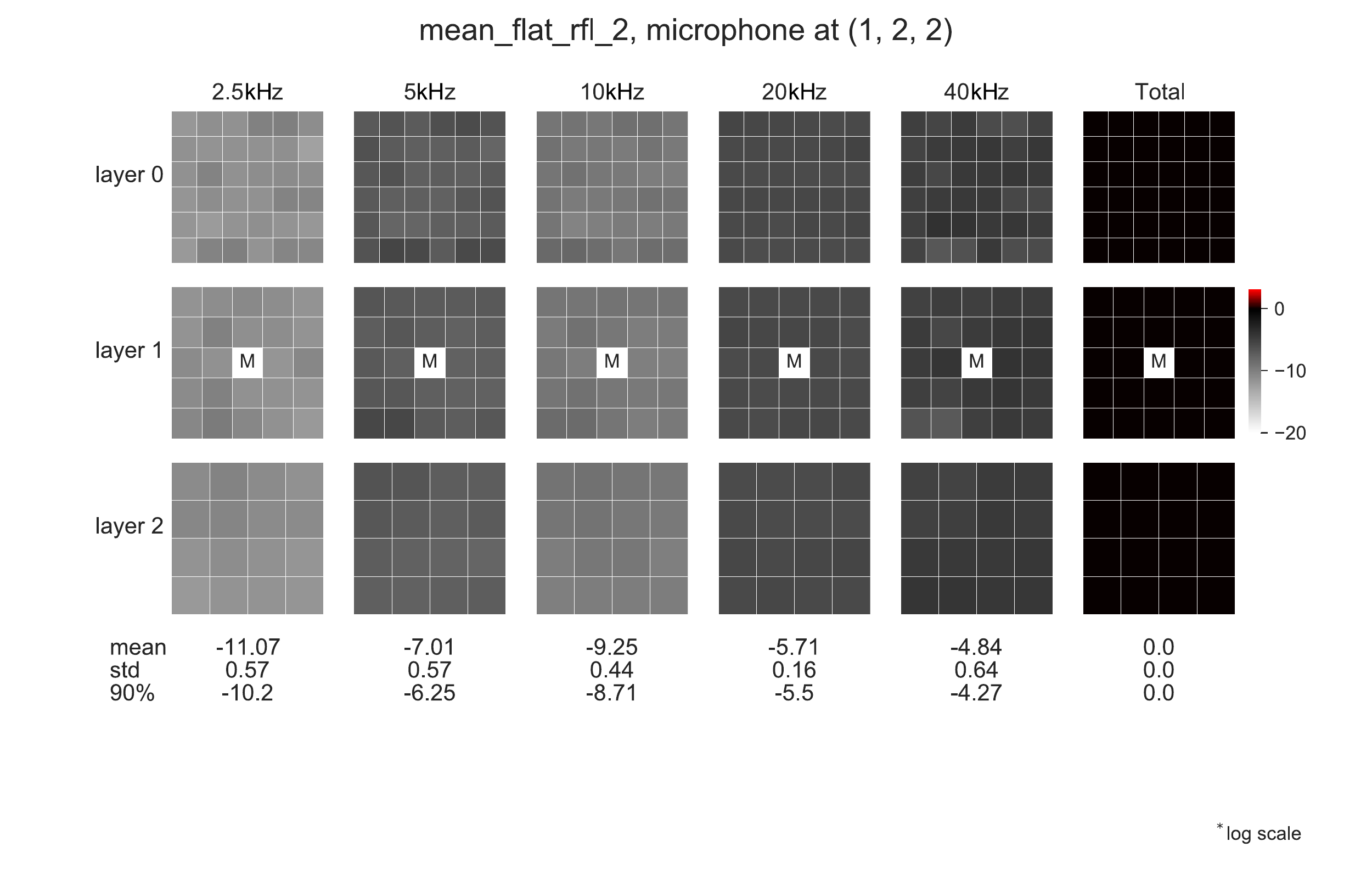}}
        \includegraphics[trim={.07\wd0} {.19\wd0} {.04\wd0} {.035\wd0}, width=\textwidth, clip]{images/mean_flat_rfl_2_122_ce_grid.pdf}
    	\caption{Flat}
    	\label{fig:flat-grid-cum-energy}		
	\end{subfigure}
	\begin{subfigure}[t]{.48\textwidth}
		\sbox0{\includegraphics{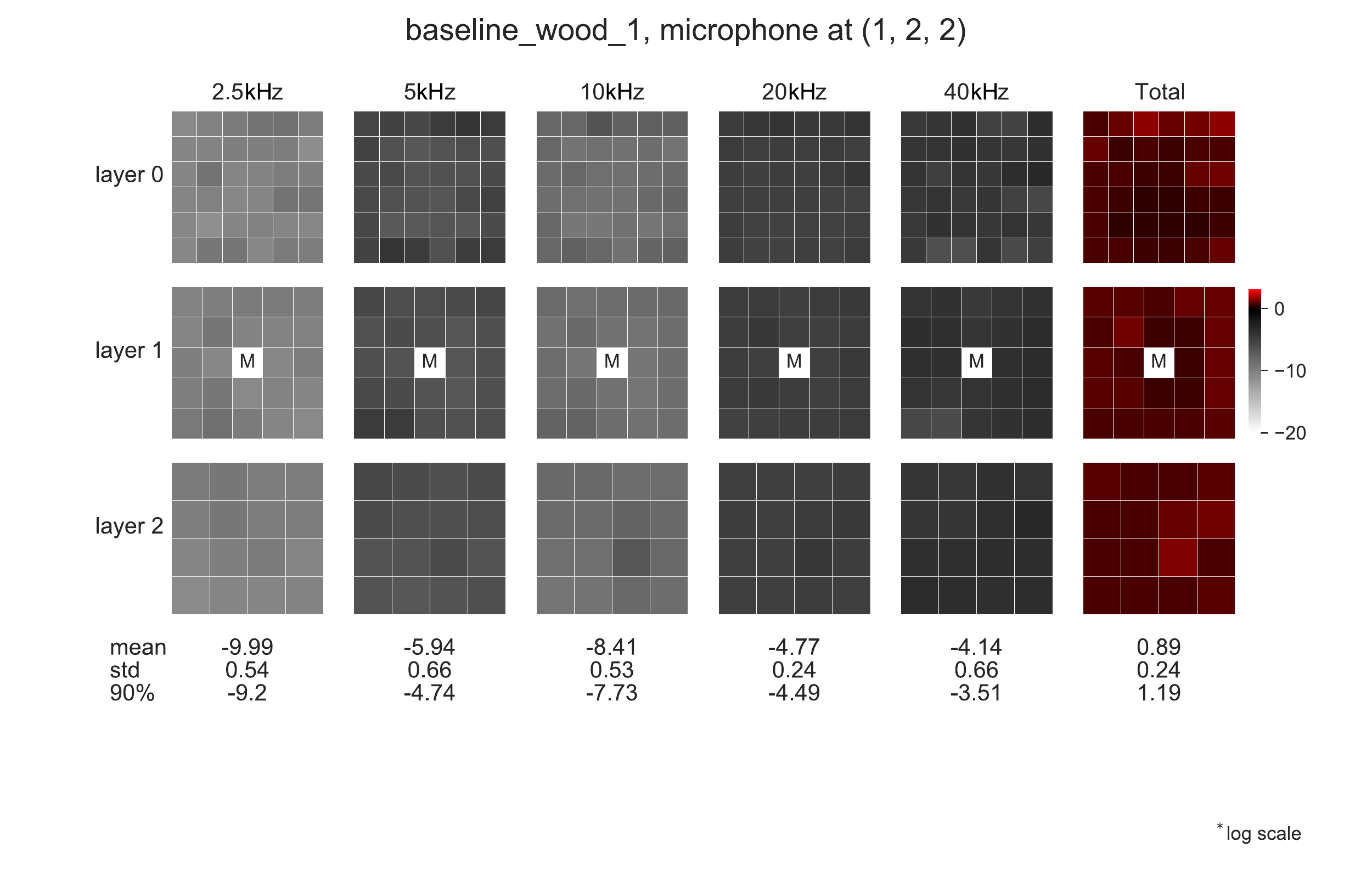}}
        \includegraphics[trim={.07\wd0} {.19\wd0} {.04\wd0} {.035\wd0}, width=\textwidth, clip]{images/baseline_wood_1_122_ce_grid.pdf}
		\caption{Wood}
		\label{fig:wood-grid-cum-energy}
	\end{subfigure}
	\begin{subfigure}[t]{.48\textwidth}
		\sbox0{\includegraphics{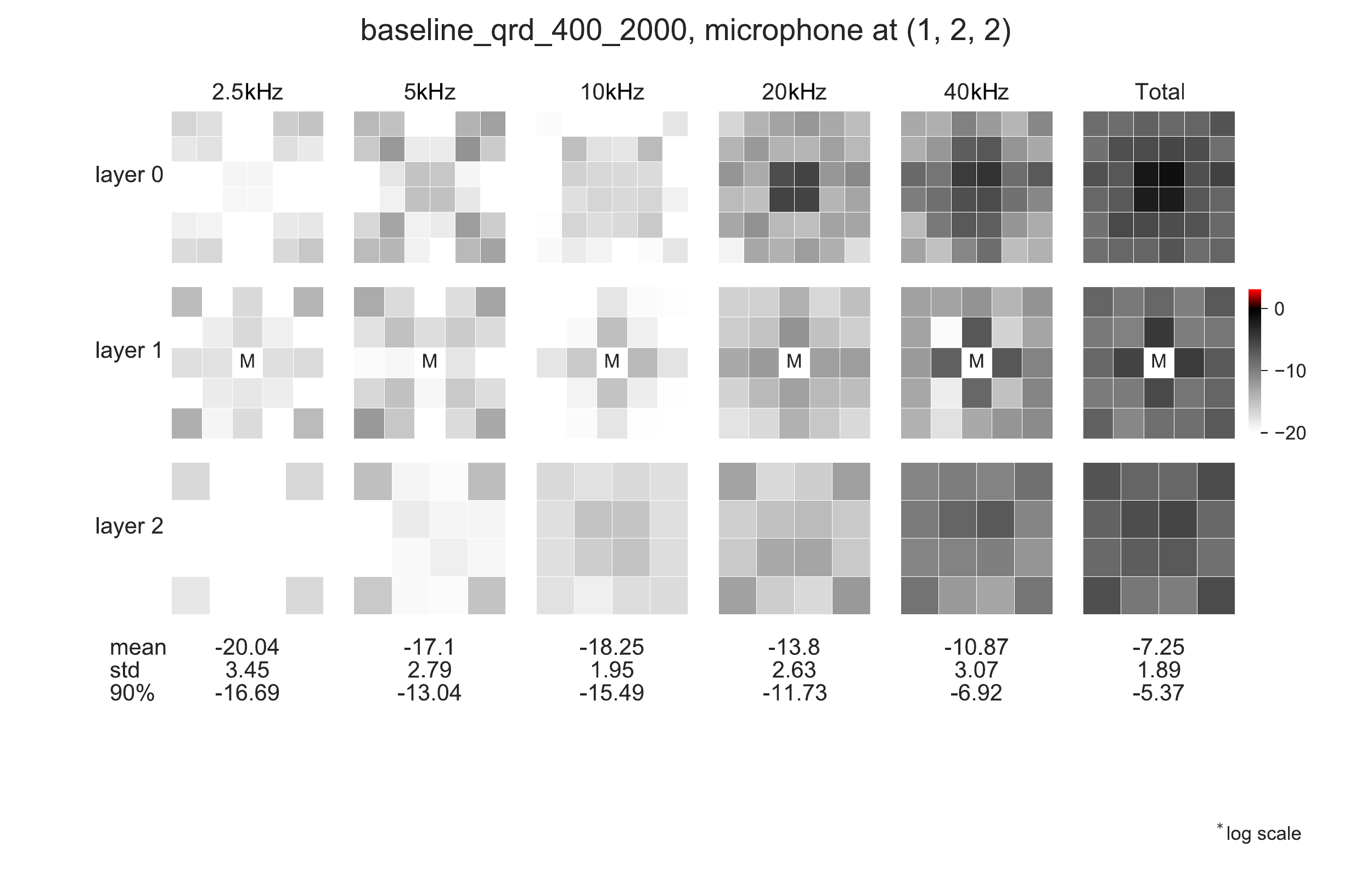}}
        \includegraphics[trim={.07\wd0} {.19\wd0} {.04\wd0} {.035\wd0}, width=\textwidth, clip]{images/baseline_qrd_400_2000_122_ce_grid.pdf}
		\caption{2D-PRD}
		\label{fig:qrd-grid-cum-energy}
	\end{subfigure}
\caption{Layered grid plot of band-separated \emph{TNCE}. Layer 0 is the closest to the surface and layer 2 the furthest away. "M" indicates the microphone's position.}
\label{fig:reference-panels-grid}
\end{figure}

For 150 measured surfaces, the \emph{TNCE} values range between 0.02 and 24.32, however the value of the 95\textsuperscript{th} percentile is 1.67. To represent those values in a compact way and to not clip high numbers, we map them on a logarithmic dB scale by applying the function \mbox{$f(x) = 10\log_{10}(x)$}. 
Figure \ref{fig:reference-panels-grid} shows all data that relate to a given microphone index at the corresponding location in the measurement grid for three of the reference panels. The \emph{TNCE} values are first mapped on the logarithmic dB scale, then converted to a color, and finally grouped based on measurement grid layers (horizontally) and filter bands (vertically). The data is always read relative to the \emph{Flat} panel: white indicates less cumulative energy (minimum -20dB) and black an equal amount (0 dB). Situations, where an amplification due to focusing occurs, are represented with red (maximum +6 dB). In this way, the high dimensional information of the panel measurements can be visually compared and evaluated (see Section \nameref{experiments}).

\begin{figure*}[!htb]
\centering
	\includegraphics[width=\textwidth]{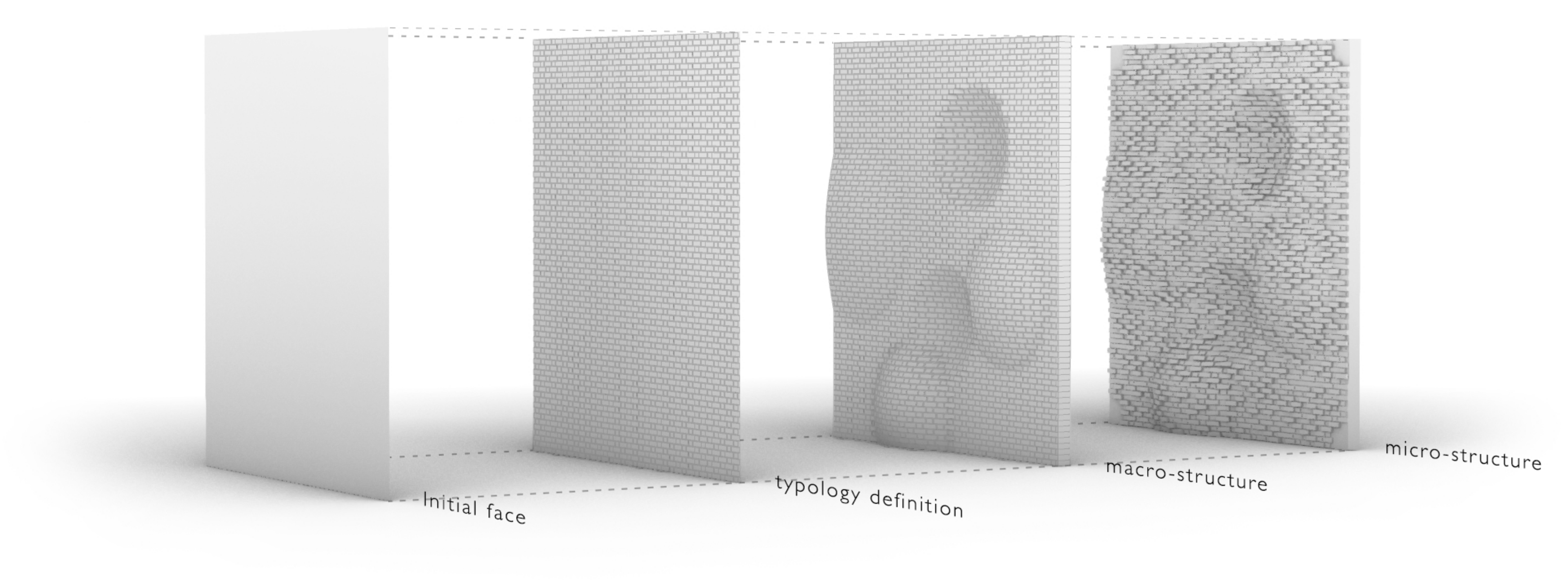}
	\caption{Geometry generation steps for a Flemish bond brick wall. From left to right: initial single-faced flat mesh, flat mesh subdivided according to typology, thickened mesh with macrostructure deformation, final mesh with microstructure deformation.}
	\label{fig:micro-meso-macro}		
\end{figure*}

\section{Diffusive surface structures}
\label{design}
Surface articulations play a significant role in the resulting acoustic response. When a sound wave is incident on a surface, the shape and size of these articulations define which frequencies will be specularly reflected and which will be scattered \cite{cox_tutorial_2006}. Diffusion is an important acoustic phenomenon that can promote spaciousness, prevent flutter echoes, and improve speech intelligibility. Although a reasonably big library of absorption coefficients for different materials is available, the same is not true for scattering coefficients \cite{coxAcousticAbsorbersDiffusers2004}.  With the goal to investigate diffuse surface properties, we generate geometric typologies stemming from architectural fabrication techniques, ensuring compatibility with past and current building systems (rubble stone walls, river rock walls, slated stone walls, brick walls). These are chosen based on their ability to diffuse sound within a broadband or a selective frequency range.
The typologies vary with the motivation to a) uncover new possibilities within the domain of acoustics, possibly integrating diffusion and absorption within one surface, and b) to diversify the acquired data-set. To ensure the latter, data acquisition and the generation of new surface geometries are performed in parallel. In this way, results from a measured panel can be used to inform the generation of new ones.

\subsection{Computational generation of diffusive surface structures} %
For each fabrication typology, the essential geometric characteristics were extracted and implemented in a geometry generation algorithm that controls the surface geometry, represented by a polygon mesh, with a set of functions.
These functions generate macro-, meso-, and microstructures based on specific criteria by applying operations such as mesh subdivision and mesh face translation. 
The macrostructures are targeting the low-frequencies, the mesostructures the mid frequencies, and the microstructures the high frequencies. For example, for a stone wall \figref{micro-meso-macro}, its general shape (depth, straight or wavy) is controlled by the macrostructure, the overall size and placement of stones by the meso-structure, and finally, the surface roughness of each stone, and the shape of the joint between them, by the microstructure. Through this modular surface generation process, panels of the same macro structure, but different micro structure, or similar combinations can be compared and analysed. 
 
Due to the limited time span of this research project, there is a limited amount of surface variations per typology that can be explored.  
At the beginning of each typology exploration, value ranges of all surface articulation parameters are defined. Then, random step sizes to sample these value ranges are chosen, and a first group with a certain number of panels is generated and produced. After this group has been measured, the acoustic data are compared against each other, and the step sizes for the next group of panels are adjusted. If the compared data are very similar, the step sizes need to be increased. If the data are significantly different, the step sizes need to be decreased. This allows for diversity in the data set while avoiding unexplored areas.

\subsection{Comparative experiments}
\label{experiments}
Comparative studies are used to investigate the relationship between geometry and resulting acoustical properties. These studies serve two main purposes: first, they help to verify certain acoustic assumptions, and thus validate the data set; and second, they serve to develop quantitative design guidelines for acoustic planners and architects, which can be used in their design workflow. For each typology, experiments are designed to test how the size, rotation, spacing, protrusion and roughness of the base element (e.g brick, stone) influence the acoustic response. This section describes three of these studies, comparing the captured acoustic data of several panels to determine how a chosen geometrical characteristic influences the acoustic performance.

\subsubsection{Brick wall joints.}
\label{sec:study-mortar}
This experiment investigates brick walls and the acoustic effect of different mortar joint heights and depths. Considering building parameters for brick walls, an average joint height ranges between 5 to 10 mm (0.5 mm to 1.0 mm in 1:10 scale). The mortar can be either flush with the surface of the bricks, or recessed by a few millimeters. Given the small size of the joint in relation to the overall surface, we only expect an influence on the high frequencies. To test this assumption, we compare three panels with the same macrostructure: two of them feature a Flemish-bond brick wall typology, with a raked joint type and an average height of 0.6 mm (\emph{0012\_1}), and 1.2 mm (\emph{0011\_1}) respectively; both with a joint depth of 1 mm. The third panel features only the macrostructure (\emph{0015\_1}), representing a brick wall with a joint height of 0.6 mm and a joint depth of 0 mm.
Table \ref{table:brick-wall-joint-height} shows the mean \emph{TNCE} of the 90\textsuperscript{th} percentile 
for each filter band. As expected, the joint height does not affect the first three filter bands (2.5, 5, 10 kHz). For the two higher ones, panel \emph{0012\_1} shows 1.04 and 1.84 dB less energy compared to the reference panel \emph{0015\_1}, and panel \emph{0011\_1} 2.6 and 3.01 dB respectively. Therefore, a small, but noticeable, reduction in the high frequencies energy can be achieved just by recessing the mortar joint and by increasing the mortar height. It is important to note that if we look at the mean \emph{TNCE} for the full spectrum, the difference is very small. Both panel \emph{0011\_1} and panel \emph{0012\_1} have very similar values and are only 0.68 and 0.46 dB respectively less than the one from panel \emph{0015\_1}.

\begin{table}[!htb]
\centering
\resizebox{\linewidth}{!}{
\begin{tabular}{cccccccc} 
\hline
Panel & Mortar & 2.5 kHz & 5 kHz & 10 kHz & 20 kHz & 40 kHz & Total \\ 
\cline{3-8}
ID & height & 90\% & 90\% & 90\% & 90\% & 90\% & 90\% \\ 
\hline\hline
0015\_1 & 0 & {\cellcolor[rgb]{0.918,0.6,0.6}}-7.66 & {\cellcolor[rgb]{0.918,0.6,0.6}}-3.66 & {\cellcolor[rgb]{0.918,0.6,0.6}}-9.18 & {\cellcolor[rgb]{0.918,0.6,0.6}}-7.02 & {\cellcolor[rgb]{0.918,0.6,0.6}}-5.54 & {\cellcolor[rgb]{0.918,0.6,0.6}}-0.4 \\ 
\hline
0012\_1 & 0.6mm & -7.8 & {\cellcolor[rgb]{0.714,0.843,0.659}}-4.14 & {\cellcolor[rgb]{0.714,0.843,0.659}}-9.58 & -8.06 & -7.38 & {\cellcolor[rgb]{0.714,0.843,0.659}}-1.02 \\ 
\hline
0011\_1 & 1.2mm & {\cellcolor[rgb]{0.714,0.843,0.659}}-7.86 & -3.89 & -9.19 & {\cellcolor[rgb]{0.714,0.843,0.659}}-9.62 & {\cellcolor[rgb]{0.714,0.843,0.659}}-8.55 & -0.86 \\
\hline
\end{tabular}
}
\captionsetup{justification=centering}
\caption{Mortar on brick walls experiment. Mean \emph{TNCE} values of the 90\textsuperscript{th} percentile per filter band. Mortar height in mm (1:10) and energy values in dB. For every frequency band, red indicates the value with the smallest difference to the reference \emph{Flat} and green the one with the highest. }
\label{table:brick-wall-joint-height}
\end{table}

\subsubsection{Macrostructure.}
\label{sec:study-macro}
\begin{figure}[htb]
\captionsetup[subfigure]{justification=centering}
\centering
	\begin{subfigure}[t]{.48\textwidth}
		\sbox0{\includegraphics{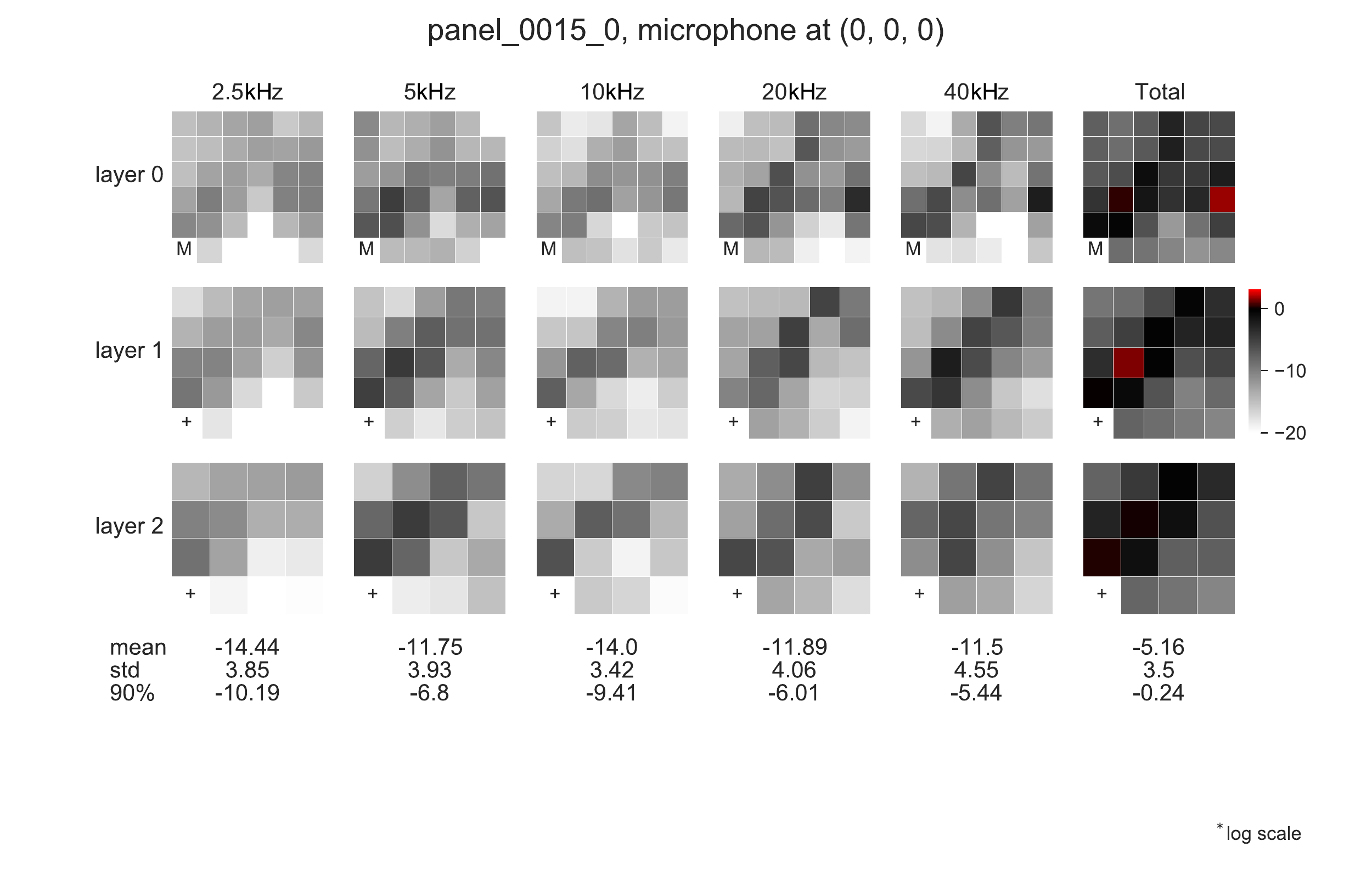}}
        \includegraphics[trim={.07\wd0} {.19\wd0} {.04\wd0} {.035\wd0}, width=\textwidth, clip]{images/panel_0015_0_000_ce_grid.pdf}
		\caption{Panel\_0015\_0}
		\label{fig:0015-grid-cum-energy}
	\end{subfigure}
	\begin{subfigure}[t]{.48\textwidth}
		\sbox0{\includegraphics{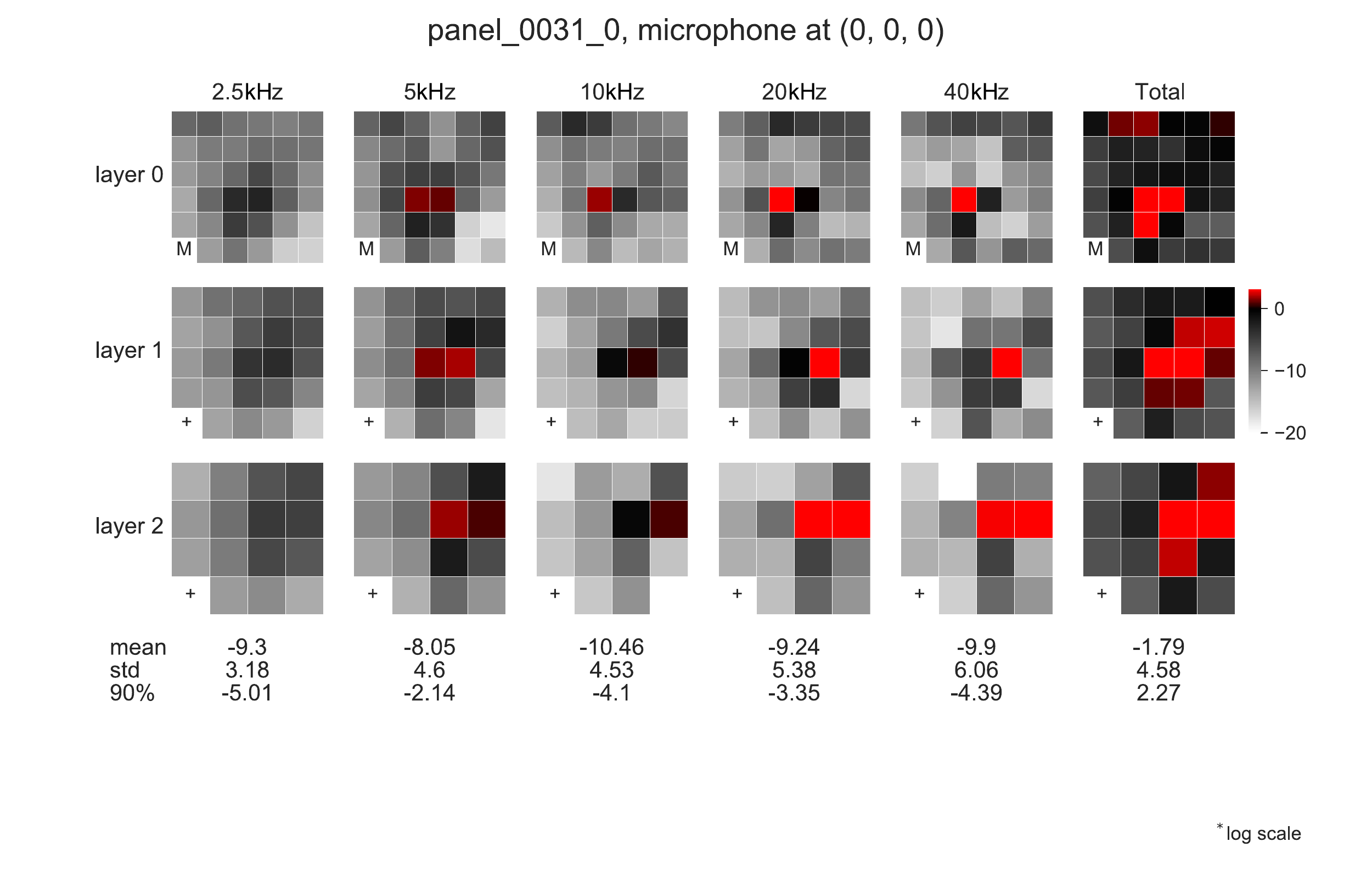}}
        \includegraphics[trim={.07\wd0} {.19\wd0} {.04\wd0} {.035\wd0}, width=\textwidth, clip]{images/panel_0031_0_000_ce_grid.pdf}
		\caption{Panel\_0031\_0}
		\label{fig:0031-grid-cum-energy}
	\end{subfigure}
\caption{\emph{TNCE} values for macrostructure comparative experiment with varying macrostructures. "M" indicates the microphone's position and "+" a point with no data.}
\label{fig:macrostructure-plot}
\end{figure}

In this experiment we compare panels \emph{Flat}, \emph{0015\_0}, and \emph{0031\_0} and focus solely on the effect of the macrostructure. Each panel has a different macrostructure, but no microstructure. Compared to \emph{Flat} (see Fig. \ref{fig:flat-grid-cum-energy}), an apparent disruption on the homogeneity of the energy distribution is visible \figref{macrostructure-plot}. Microphone-speaker combinations where their Fresnel zone falls in a convex shape exhibit less energy. Contrary, combinations in which their Fresnel zone falls in a concave part of the surface exhibit increased energy due to the focusing effect (see red squares in Fig. \ref{fig:0031-grid-cum-energy}). When comparing two panels that share the same macrostructure but have different microstructures \figref{stones-bricks-plot}, the cumulative energy plots show that the macrostructure influences all frequency bands and the microstructures start having an influence only after 10 kHz.


\subsubsection{Stone vs. brick walls.}
\label{sec:study-stones-bricks}
\begin{figure}[htb]
\centering
	\includegraphics[width=\linewidth]{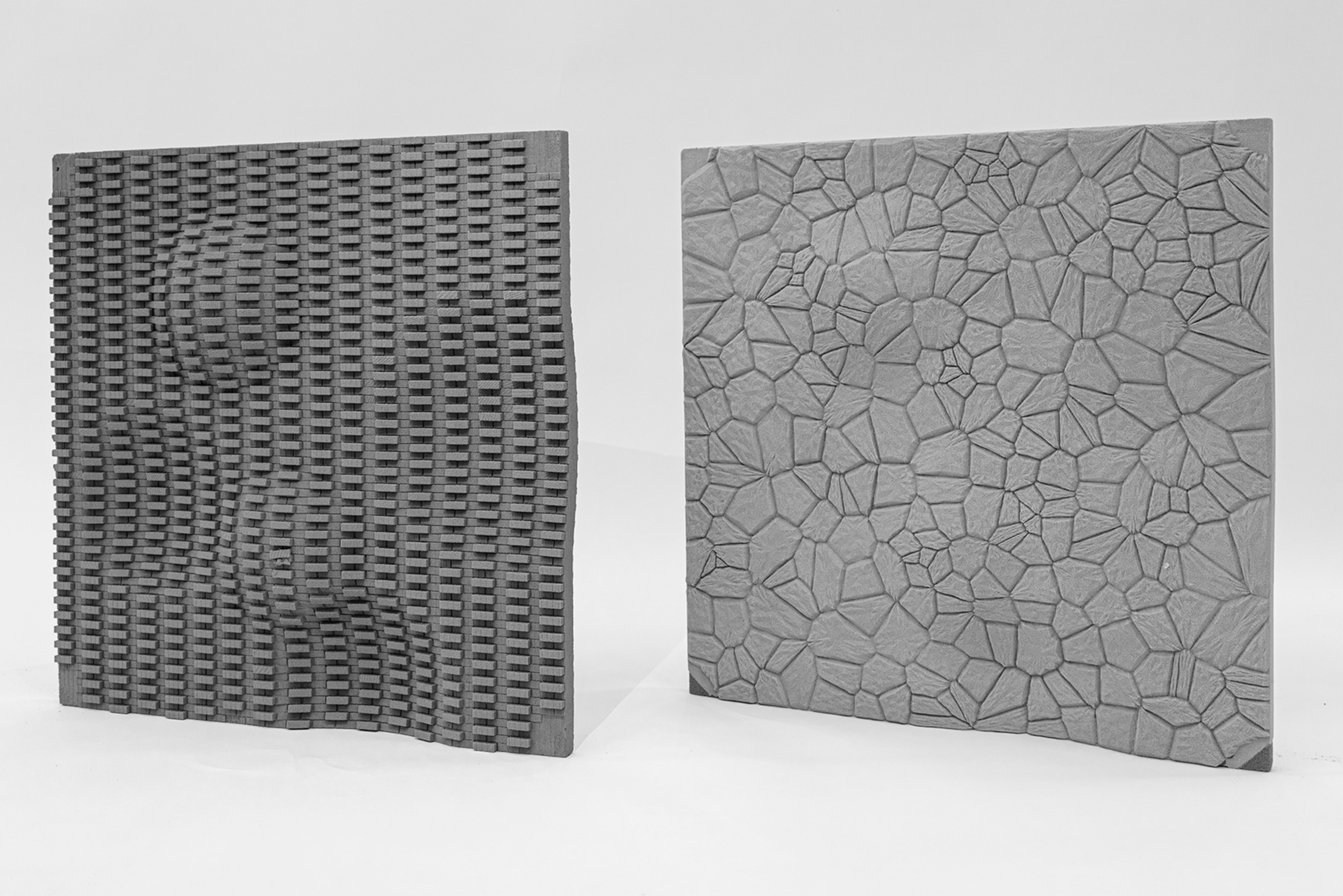}
	\caption{3D printed and coated panels. Left: Surface \emph{0013\_1} from the brick typology. Right: Surface \emph{0005\_1} from the polygonal rubble stone wall typology.}
	\label{fig:3d-printed-panels}		
\end{figure}

This experiment aims to determine whether stone walls or brick walls are better in diffusing sound. We generated and measured 46 stone and 92 brick walls. Our analysis shows that brick walls diffuse sound more consistently. Polygonal rubble stone walls generally diffuse less energy in the lower frequencies, but the results were inconclusive for the mid and high frequencies. To illustrate the findings we present two extreme cases (see Fig. \ref{fig:3d-printed-panels}).

\begin{figure}[!htb]
\captionsetup[subfigure]{justification=centering}
\centering
    \begin{subfigure}[t]{.48\textwidth}
       \sbox0{\includegraphics{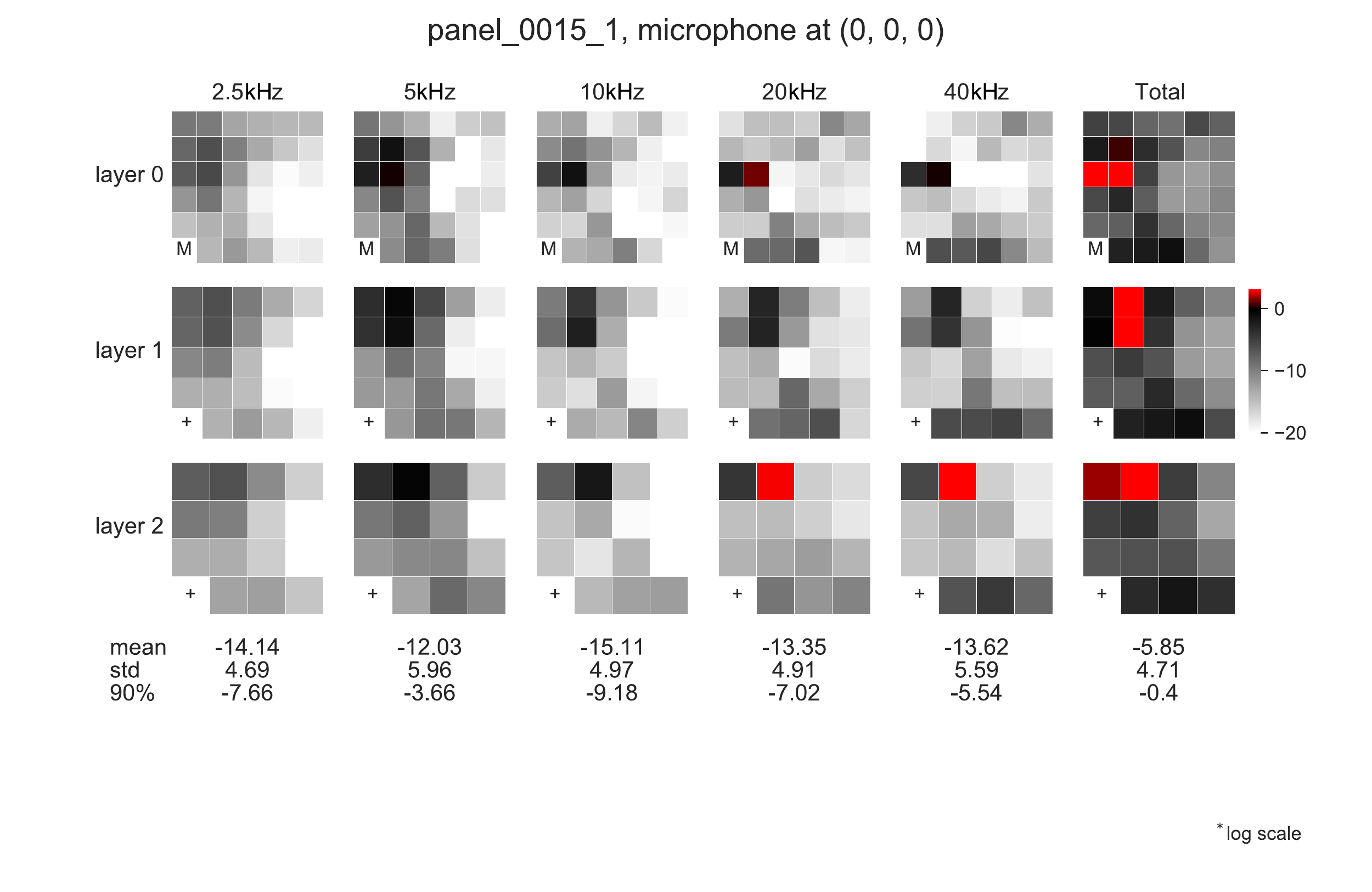}}%
        \includegraphics[trim={.07\wd0} {.19\wd0} {.04\wd0} {.035\wd0}, width=\textwidth, clip]{images/panel_0015_1_000_ce_grid.pdf}
    	\caption{Panel \emph{0015\_1}}
    	\label{fig:macro-grid-cum-energy}		
	\end{subfigure}
 	\hfill
	\begin{subfigure}[t]{.48\textwidth}
		\sbox0{\includegraphics{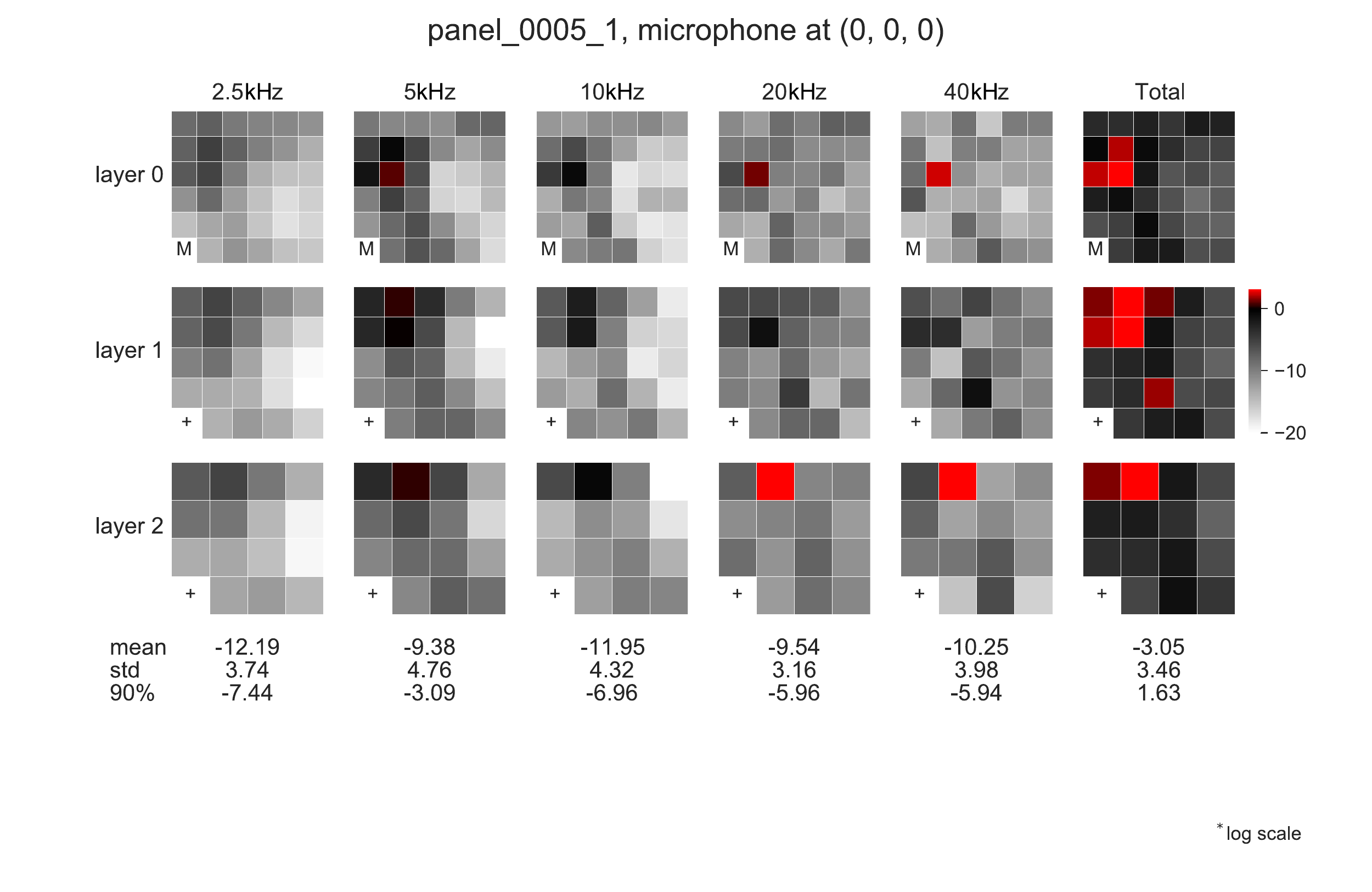}}%
        \includegraphics[trim={.07\wd0} {.19\wd0} {.04\wd0} {.035\wd0}, width=\textwidth, clip]{images/panel_0005_1_000_ce_grid.pdf}
		\caption{Panel \emph{0005\_1}}
		\label{fig:0005-1-grid-cum-energy}
	\end{subfigure}
 	\hfill
	\begin{subfigure}[t]{.48\textwidth}
		\sbox0{\includegraphics{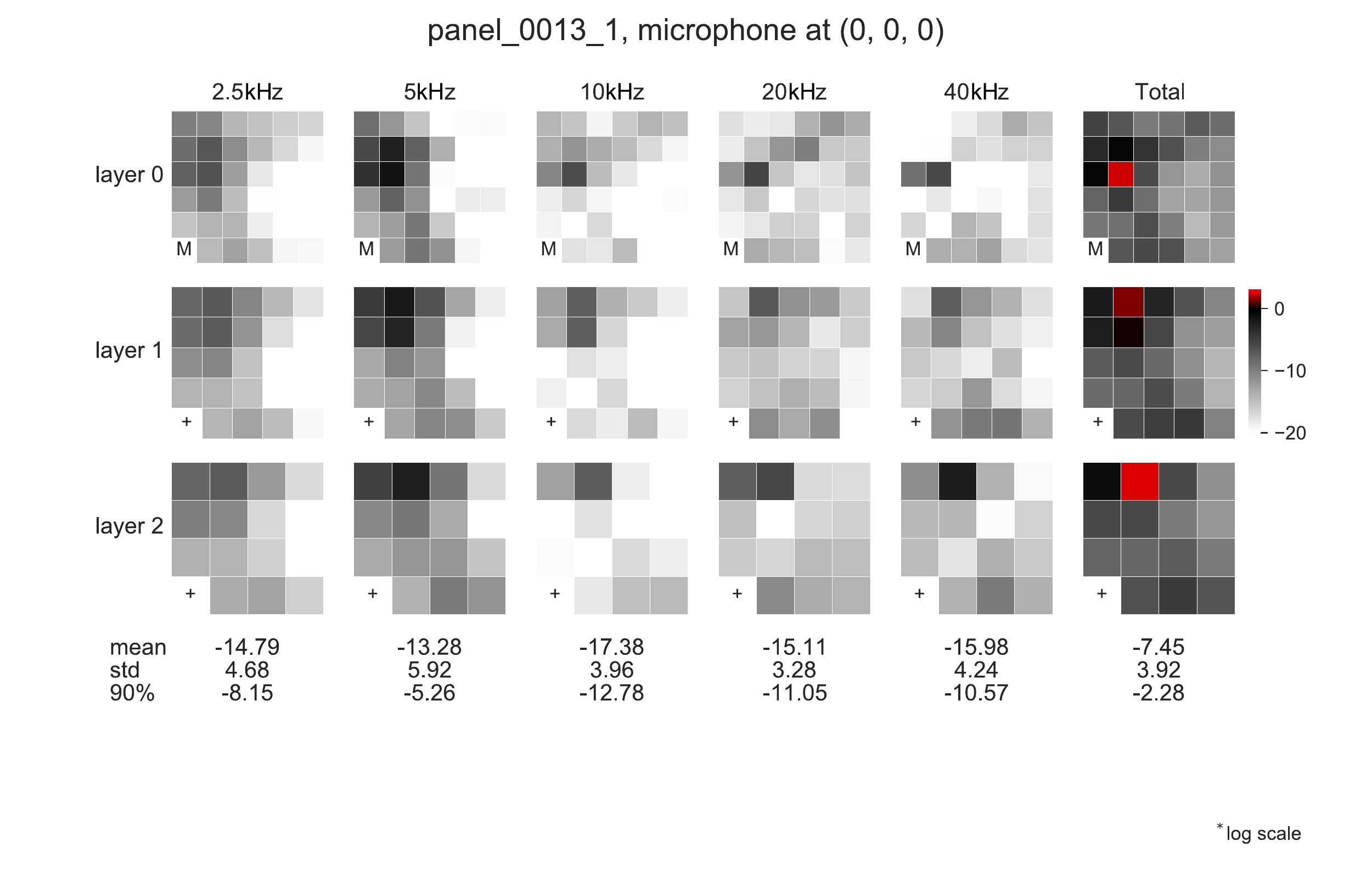}}%
       \includegraphics[trim={.07\wd0} {.19\wd0} {.04\wd0} {.035\wd0}, width=\textwidth, clip]{images/panel_0013_1_000_ce_grid.pdf}
		\caption{Panel \emph{0013\_1}}
		\label{fig:0013-1-grid-cum-energy}
	\end{subfigure}
\caption{\emph{TNCE} values for stone vs brick walls - experiment. (\ref{fig:macro-grid-cum-energy}) Panel with only the macrostructure, (\ref{fig:0005-1-grid-cum-energy}) Stone wall typology panel with the same macrostructure as panel \emph{0015\_1}, (\ref{fig:0013-1-grid-cum-energy}) Brick wall typology panel with the same macrostructure as panel \emph{0015\_1}}
\label{fig:stones-bricks-plot}
\end{figure}

Panel \emph{0005\_1} is from the polygonal rubble stone wall typology. It features, on average, nine stones per square meter, a joint width between 20-30 mm, a joint depth between 50-80 mm, and a stone surface roughness of \rpm30 mm (numbers in 1:1). Panel \emph{0013\_1} is from the brick typology and resembles a standard stretcher-bond brick wall. It features standard bricks measuring 215x65x102.5 mm (WxHxD) and a raked joint around 15 mm wide (\rpm1 mm) and 10 mm deep (\rpm1 mm) 
(numbers in 1:1). 
Both panels \emph{0005\_1} and \emph{0013\_1} share the same macrostructure with panel \emph{0015\_1}.
Compared to panel \emph{0015\_1} (Fig. \ref{fig:macro-grid-cum-energy}), panel \emph{0005\_1} (Fig. \ref{fig:0005-1-grid-cum-energy}) exhibits higher \emph{TNCE} values across all filter bands (see Table \ref{table:stone-vs-brick}), with the exception of 40 kHz, but only by 0.4 dB. On the contrary, panel \emph{0013\_1} (Fig. \ref{fig:0013-1-grid-cum-energy}) exhibits less cumulative energy across all filter bands. The difference is smaller in the two lower filter bands (-0.49, -1.6 dB), in which the effect of the macrostructure is more dominant, but more present in the upper three bands (-3.6, -4.03, -5.03 dB). In comparison to panel \emph{0015\_1}, the \emph{TNCE} of panel \emph{0005\_1} is higher by 2.03 dB, and of panel \emph{0013\_1} lower by -1.88 dB.

\begin{table}[htb]
\centering
\resizebox{\linewidth}{!}{
\begin{tabular}{ccccccc} 
\hline
Panel & 2.5 kHz & 5 kHz & 10 kHz & 20 kHz & 40 kHz & Total \\ 
\cline{2-7}
ID & 90\% & 90\% & 90\% & 90\% & 90\% & 90\% \\ 
\hline\hline
0015\_1 & -7.66 & -3.66 & -9.18 & -7.02 & {\cellcolor[rgb]{0.918,0.6,0.6}}-5.54 & -0.4 \\ 
\hline
0005\_1 & {\cellcolor[rgb]{0.918,0.6,0.6}}-7.44 & {\cellcolor[rgb]{0.918,0.6,0.6}}-3.09 & {\cellcolor[rgb]{0.918,0.6,0.6}}-6.96 & {\cellcolor[rgb]{0.918,0.6,0.6}}-5.96 & -5.94 & {\cellcolor[rgb]{0.918,0.6,0.6}}1.63 \\ 
\hline
0013\_1 & {\cellcolor[rgb]{0.714,0.843,0.659}}-8.15 & {\cellcolor[rgb]{0.714,0.843,0.659}}-5.26 & {\cellcolor[rgb]{0.714,0.843,0.659}}-12.78 & {\cellcolor[rgb]{0.714,0.843,0.659}}-11.05 & {\cellcolor[rgb]{0.714,0.843,0.659}}-10.57 & {\cellcolor[rgb]{0.714,0.843,0.659}}-2.28 \\
\hline
\end{tabular}
}
\captionsetup{justification=centering}
\caption{Stone vs. brick walls. Mean TNCE of the 90\textsuperscript{th} percentile per filter band. All values are relative to the reference \emph{Flat}.}
\label{table:stone-vs-brick}
\end{table}

\section{Conclusions and future work}
\label{conclusion}

In this paper, we presented a novel approach to study the mutual relationship between diffusive surface structures and their acoustic performance through data science methods. We described the post-processing of the measured data, the evaluated indicators and showed that they can be used for the quantitative assessment of different surface structures; thus providing a valid evaluation system.

By the end of this research project we target to measure 350 acoustic surfaces; this amounts to approximately 1 million impulse responses in total. To create high diversity, the data set is continuously shaped through analysis and subsequent surface generation. This approach should enable the future ML model to generalize as well as possible. Analytical tools, for example PCA \cite{karl_pearson_frs_liii_1901} and PCC \cite{benesty_pearson_2009}, will be evaluated to identify the most important geometrical characteristics that influence certain acoustic responses. Self-organising maps (SOM) \cite{kohonen_self-organized_1982} are also tested to cluster the surfaces based on geometrical (fabrication typology) and acoustical characteristics (i.e. absorption, scattering). 
This will help to identify unexplored areas in the design and data space and test new hypotheses that emerge during analysis.

Currently, we can only speculate on the output and accuracy of the ML system. However it is foreseeable that the limited number of data set samples will be critical. This limitation can be mitigated by leveraging the large number of impulse responses present in each sample and by the data augmentation naturally emerging from the setup symmetries. From our preliminary testing on 100 acoustic surfaces, we believe that an exact prediction of the impulse response is likely impossible. Hence, we will focus our efforts on predicting the compressed information obtained from the post-processing step presented in this paper. Our preliminary ML architecture is able to predict the energy reflected (more precisely the TNCE) in every measured position for geometries that present similarities with the training set.

Our future ML model shall be used as a fast acoustic evaluation tool for diffusive surfaces, which facilitates acoustic driven form-finding in early design phases. Together with the developed design guidelines for certain fabrication typologies, this will enable more acoustic aware designs, thus bringing acoustics closer to the architectural practice.

\subsection{Copyright}
Copyright \copyright\ \volumeyear\ SAGE Publications Ltd,
1 Oliver's Yard, 55 City Road, London, EC1Y~1SP, UK. All
rights reserved.

\begin{acks}
This research is jointly supported by the Swiss Data Science Center and the Chair of Architecture and Digital Fabrication, ETH Zurich. The authors would like to thank Michael Lyrenmann and Philippe Fleischmann from the NCCR Digital Fabrication, ETH Zurich, for their help with building the multi-robotic measurement setup, Dr. Mariana A. Popescu for her assistance producing the 3D knitted sleeves for the robots, and Anton Johansson for his help in the measurement process. Furthermore, Dr. Mathias Bernhard, Patrick Bedarf, and Pietro Odaglia from the Chair of Digital Building Technologies, ETH Zurich for their support with the in-house 3D printer and Alessandro Tellini from Raplab, ETH Zurich for his help and support using the digital workshop, and finally, a sincere thank you to Dr. Lauren Vasey for her diligent proofreading of this manuscript.
\end{acks}

\section{APPENDIX}
\subsection{Impulse response post-processing details}
\label{appendix:impulse_response}
\begin{figure}[htb]
\centering
	\includegraphics[width=\linewidth]{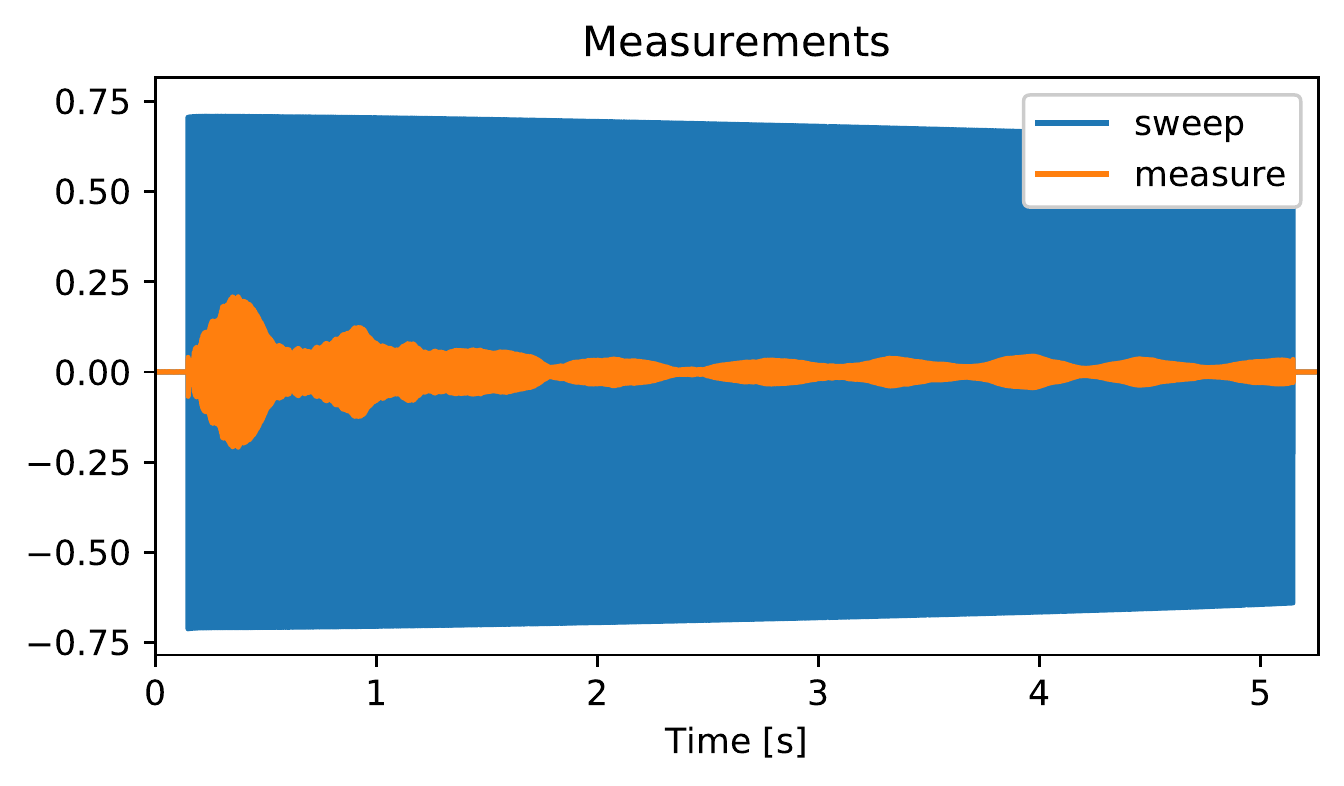}
	\includegraphics[width=\linewidth]{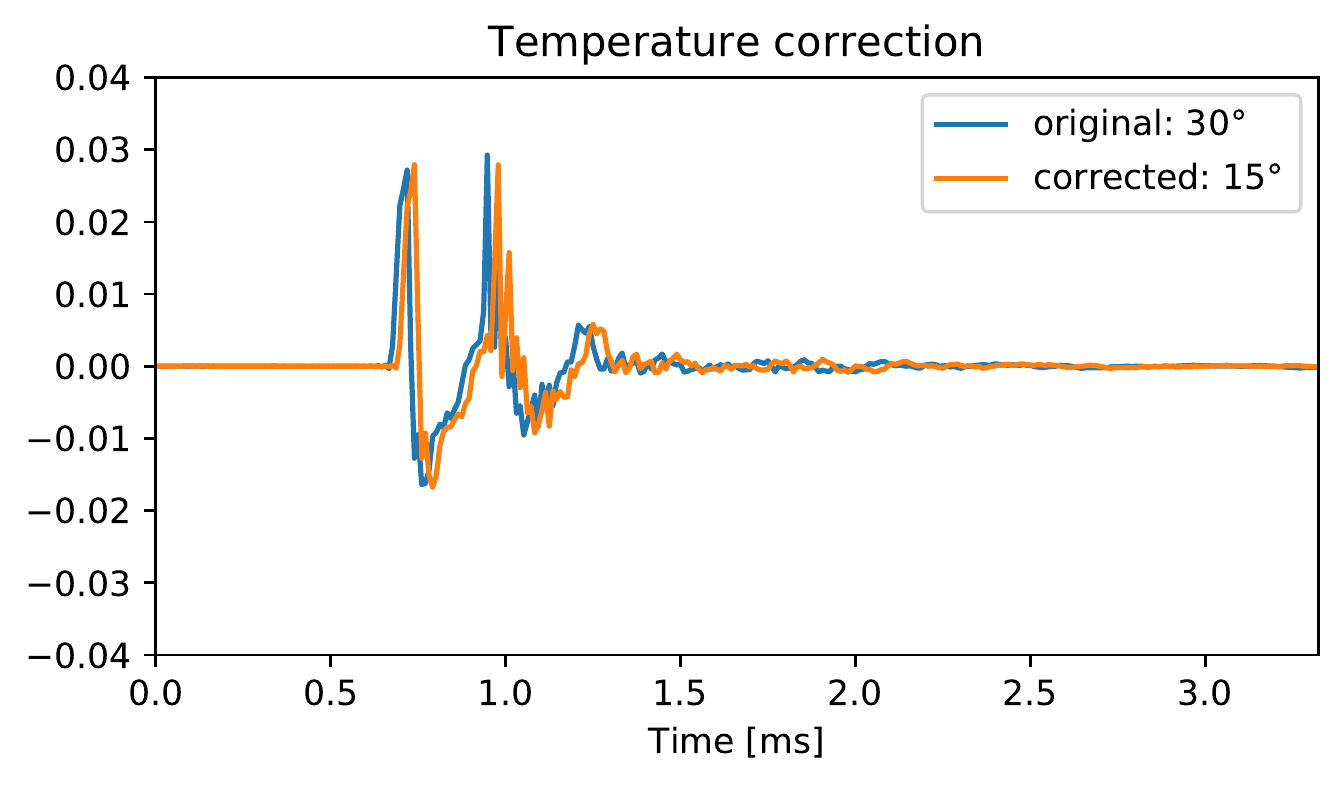}
	\includegraphics[width=\linewidth]{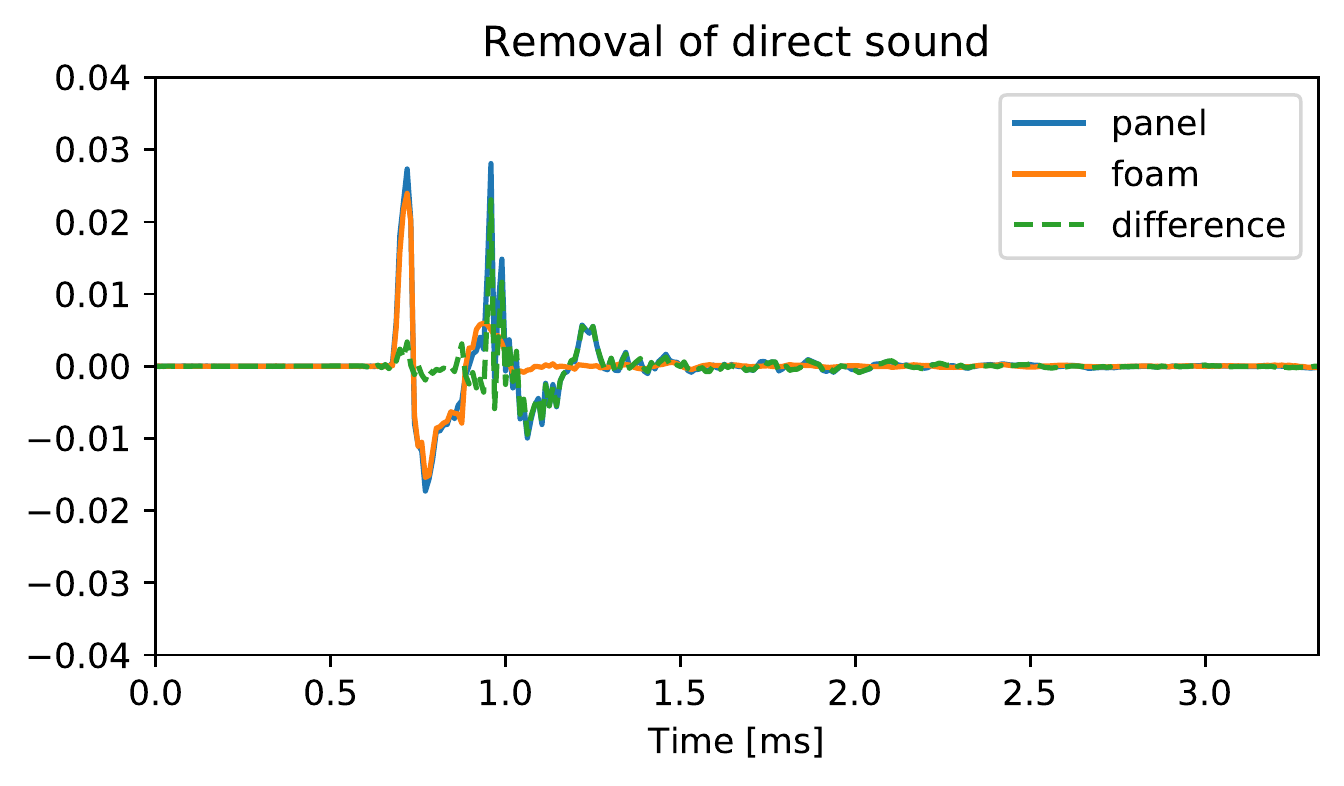}
	\caption{Impulse response post-processing. Top: measurements. Middle: after deconvolution, temperature correction is applied. We used 30$\degree C$ and 15$\degree C$ to emphasize the effect of resampling.  Bottom: eventually direct sound is removed.}
	\label{fig:impulse_post_process}		
\end{figure}

The post-processing of the impulse response consists of the following 3 steps which are illustrated in Figure~\ref{fig:impulse_post_process}:
\paragraph{1. Deconvolution.} The deconvolution operation is carried out using a simple division in the Fourier domain. Given $\hat{x} = \mathcal{F}x $ the Fourier transform of $x$ and $x = \mathcal{F}^{-1} \hat{x}$ its inverse operation, the deconvolution of the signal $x$ with the sweep $s$ is given by
\begin{equation*}
x_d = \mathcal{F}^{-1}\left( \mathcal{F}x / \mathcal{F}s \right),
\end{equation*}
where $s$ is the sweep and the division is performed elementwise. Note that $\mathcal{F}s$ is never close to 0 because the sweep contains all frequencies.

\paragraph{2. Temperature correction.}
To adjust for the room temperature change, we estimate the speed of sound at temperature $T$ (in $\degree C$ )
\begin{equation*}
  c = c_0 \sqrt{1 + (T / 273.15)},
\end{equation*}
where $c_0$ is the temperature at 0$\degree C$~\cite{cramer1993variation}. The impulse response is then resampled at the frequency $c/c_{ref} f_s$, where $c_{ref}$ is the speed of sound at 20$\degree C$ and $f_s=96kHz$ the sampling frequency.
We use the polyphase filtering method (``resample\_poly``) from the SciPy python package).

\paragraph{3. Removal of direct sound.}
The direct sound removal is performed by subtracting the impulse response of the absence of a wall (an absorbent foam inserted instead of the panel, see Table \ref{tab:baseline}, \emph{Foam}).

\bibliographystyle{SageV}
\bibliography{gkr_ddad} 
\end{document}